\documentclass[12pt]{article}
\input{definition}

\begin{document}
	\begin{center}
		{\bf\Large  Model-Free, Monotone Invariant and Computationally Efficient Feature Screening with Data-adaptive Threshold}
		\\\vskip 0.3cm
		LINSUI DENG and YILIN ZHANG
		\\\vskip0.3cm
		\textit{ 
			Renmin University of China}
		\begin{singlespace}
			\footnotetext[1]{Linsui Deng (Email: denglinsui@ruc.edu.cn)  and Yilin Zhang (Email: yzhang97@ruc.edu.cn), Center for Applied Statistics and  	Institute of Statistics and Big Data,    	Renmin University of China, 
			Beijing 100872, China. Yilin Zhang is corresponding author. Zhang's research  is supported by the Fundamental Research Fund  for Central Universities  and  the Research Fund  of Renmin University of China (21XNH157).}
		\end{singlespace}
		\today
	\end{center}

\begin{singlespace}
	\begin{abstract}
 Feature screening for ultrahigh-dimension, in general, proceeds with two essential steps. The first step is measuring and ranking the marginal dependence between response and covariates, and the second is determining the threshold.
We develop a new screening procedure, called SIT-BY procedure, that possesses appealing statistical properties in both steps. 
By employing sliced independence estimates in the measuring and ranking stage, our proposed procedure requires no model assumptions, remains invariant to monotone transformation, and achieves almost linear computation complexity. 
Inspired by false discovery rate (FDR) control procedures, we offer a data-adaptive threshold benefit from the asymptotic normality of test statistics. 
Under moderate conditions, we demonstrate that our procedure can asymptotically control the FDR while maintaining the sure screening property. 
We investigate the finite sample performance of our proposed procedure via extensive simulations and an application to genome-wide dataset. 
\end{abstract}
\end{singlespace}
	

\newpage
\csection{Introduction}
Ultrahigh-dimensional data analysis has attracted much attention due to its widespread applications in fields such as biological science, computer and information science, economics, and astronomy \citep{Fan2020Stat}. 
To exclude redundant features in the ultrahigh-dimensional case, an effective paradigm, called sure independence screening, was proposed in \cite{fan2008sure}.
This procedure has aroused extensive concern in the statistical field for its notable sure screening property, that is selecting all important features with probability one. 
In general, a feature screening procedure proceeds with two steps.
The first step is measuring the marginal dependence between the response and covariates and ranking the covariates according to their dependence.  
The second is identifying a threshold to separate the active covariates and inactive ones.

There have been numerous studies focusing on the measuring and ranking step. Classical model-based screening methods consider widespread cases, including linear model \citep{fan2008sure} and its variants \citep{hall2009using, fan2010sure, fan2011nonparametric}. 
In recent years, an attractive measuring and ranking procedure should possess the following three properties. 
The first is model-free, that is 
allowing arbitrarily nonlinear dependence between the response and covariates. 
In contrast, imposing a parametric form runs the risk of model misspecification, for there is usually little prior knowledge about the actual dependence structure \citep{zhu2011model, li2012feature, mai2015fused}. 
The second is monotone invariant, which means that the dependence measure remains unchanged when applying any monotone transformation to response and covariates. 
This property ensures yielding consistent conclusions in ranking covariates, and thus allows both response and covariates to be heterogeneous or heavy-tailed \citep{zhou2018ModelfreeFeatureScreening,kong2019CompositeCoefficientDetermination}.
The third is numerically efficient. 
For the screening procedure going through each covariate, the computational complexity of nonlinear measure estimates is expected to reach the optimal, i.e., linear or nearly linear in sample size.

All the aforementioned properties are fulfilled by sliced independence estimates proposed in \cite{zhang2021SlicedIndependenceTest}, which enhance the power of well-known rank-based estimates \citep{chatterjee2020NewCoefficientCorrelation}
and accelerate the kernel smoothing-based estimates \citep{kong2019CompositeCoefficientDetermination}.
Therefore, we highly advocate applying the sliced independence estimates in screening and ranking step. 
In addition, the sliced independence estimates converge to normal distribution, which largely facilitates choosing threshold.


Threshold determination
is of equal importance in feature screening. An ideal threshold should ensure sure screening property, while simultaneously including as few redundant covariates as possible \citep{Guo2021}.  
The hard thresholding rule, widely adopted in screening methods, selects covariates with the $d=\lfloor n/\log n\rfloor$ largest marginal dependence measures, where $n$ is the sample size \citep{fan2008sure}. 
Still, due to fully determination by the sample size, it can face the challenge of being conservative when the sample size is small and being aggressive when the sample size is large \citep{Guo2021}. 
\cite{zhu2011model} accounted for the effect of signals and offered a soft threshold by mimicking the performance of inactive features via artificial auxiliary features. 
However, it remains unclear whether the chosen threshold 
still guarantees the sure screening property.
Recently, choosing a data-adaptive threshold controlling false discovery rate (FDR, \citealp{Benjamini1995}) has become a hot research topic. 
In this scope, the main purpose is to control the proportion of inactive covariates among the identified covariates.  
\cite{Liu2022}, \cite{Guo2021} and \cite{tong2022model} constructed marginal  measures symmetric to zero under independence case.
Their theoretical guarantees adapt from the knockoff framework \citep{Barber2015, Candes2018}. Surprisingly, the data-adaptive threshold simultaneously realizes the sure screening property in some situations \citep{Liu2022, Guo2021}. 

However, ignoring the marginal distributions, if available, might deteriorate the detection efficiency due to the information loss.
The Benjanimini-Hochberg procedure (BH procedure) is anticipated to perform better because it uses the information about distribution. 
The BH procedure imposes requirements on the dependency among the marginal utility measures, which are nearly impossible to verify. 
The Benjamin-Yekutieli procedure (BY procedure) rigorously controls FDR as long as the distribution of marginal utility measures of inactive covariates is given \citep{Benjamini2001}. 
Given that the sliced independence estimates are all asymptotically normally distributed \citep{zhang2021SlicedIndependenceTest}, the BY procedure is completely suitable to obtain a data-adaptive threshold. 
More importantly, we can further build the sure screening property given the threshold from the BY procedure.

The remainder of this paper is organized as follows. 
In Section \ref{sec:sis}, we review the sliced independence test and propose our screening procedure, which is efficient, model-free, and monotone invariant. We establish the convergence rate, sure screening property, and rank consistency property under mild conditions. 
In Section \ref{sec:sisfdr}, we provide a data-adaptive threshold for realizing sure screening property and controlling FDR simultaneously and examine the corresponding theoretical properties. 
In Section \ref{sec:simu}, we evaluate the performance of the proposed screening procedure in terms of sure screening property and FDR control through numerical simulations. We further apply the proposed procedure to a rat eye gene expression dataset to demonstrate its effectiveness. 
Section~\ref{sec:conclusion} briefly summarizes the paper.  
All technique proofs are relegated in Supplementary Material.

\csection{An Model-Free, Monotone Invariant and Computationally Efficient Feature Screening Procedure}\label{sec:sis}

\csubsection{A general framework}\label{sec:framework}
To begin with, we formulate a general framework for feature screening problem. 
Suppose $Y\in\mR^1$ is a univariate response and $\x = (X_1,\ldots, X_p)\trans\in\mR^p$ is a $p$-vector of covariates.  
Our goal is to identify the covariates informative to the response. To be precise, we write $\x = (\x_\calA\trans, \x_\calI\trans)\trans$, where $\x_\calA$ stands for the active covariates which are predictive for the response, and $\x_\calI$ stands for the inactive covariates which do not provide any additional information for predicting $Y$ once $\x_\calA$ is given. 
In symbols,
\beqr\label{model}
Y \ \hDash\  \x \mid \ \x_\calA,
\eeqr
where ``$\hDash$" stands for ``statistical independence". Model   (\ref{model}) is equivalent to stating that $Y  \hDash  \x_\calI \mid \ \x_\calA$. In order to simultaneously enhance predictability and interpretability of subsequent modeling, it is important to retain $\x_\calA$ and remove as many inactive covariates in $\x_\calI$ as possible. 

Model \eqref{model} has three important implications. 
First, Model \eqref{model} allows $Y$ to depend upon $\x_\calA$ in an arbitrarily nonlinear fashion, rather than in a pre-specified parametric form, which is also called model-free. 
Second, Model \eqref{model} is invariant to monotone transformations. To be precise, let $U = m_{0}(Y)$, $\mathbf{v} = (V_1,\ldots, V_p)\trans$ and $V_k = m_k(X_k)$, where $m_k$ is a strictly monotone transformation function, for $k = 0,1,\ldots, p$. Then, Model (\ref{model})  is equivalent to  stating that $U \ \hDash\  \mathbf{v}  \mid \ \mathbf{v} _\calA$. 
Third, Model \eqref{model} takes all $p$ dimensional variables into consideration. The computational complexity will take the form of $O\{pC(n)\}$, where $C(n)$ is expected to be the almost linear function of sample size $n$. An appealing feature screening procedure is anticipated to satisfy these implications.

\csubsection{Sliced independence test}\label{sec:sit}
In this section, we introduce the sliced independence estimates proposed by \cite{zhang2021SlicedIndependenceTest}, which can fulfill all implications in Section~\ref{sec:framework}. 
Consider two univariate random variables, $X$ and $Y$, and we measure their nonlinear dependence through 
a nonlinear version of R squared coefficient
\beqr\label{definition:dependencemeasure}
\calS(X, Y)  &=&  1 - \int E\Big[\var\{1(Y\geq t)\mid X\}\Big] \mathrm{d}\mu(t)\Bigg/\int\var\{1(Y\geq t)\} \mathrm{d}\mu(t),
\eeqr 
where $\omega(t)$ and $\mu(t)$ are the probability mass/density and cumulative distribution of $T$, a univariate random variable independent to $X$ and $Y$. 
Its support, denoted as $\textrm{supp}(T)\defby \{t: \omega(t)>0 \}$, satisfies that $\textrm{supp}(Y)\subseteq \textrm{supp}(T)$.

For its appealing properties at the population level, this dependence metric and its variants have been discussed in  \cite{dette2013CopulaBasedNonparametricMeasure}, \cite{kong2019CompositeCoefficientDetermination} and \cite{chatterjee2020NewCoefficientCorrelation}. 
\cite{zhang2021SlicedIndependenceTest} stated that this dependence metric possesses  the zero equivalence property, that is, $\calS(X, Y) =0$ if and only if $X$ and $Y$ are independent, and $\calS(X, Y)  =1$ if and only if $Y$ is measurable function of $X$. More importantly, this metric has monotone invariance for both $X$ and $Y$. This means that $\calS(X, Y)$ remains unchanged, whether $X$ or $Y$ is transformed by strictly monotone function.

To estimate \eqref{definition:dependencemeasure}, \cite{zhang2021SlicedIndependenceTest} introduced the sliced independence estimates. 
Suppose $\{(X_i, Y_i), i = 1,\ldots, n\}$ are independent random sample, following the same distribution as $(X, Y)$. 
The first step is ordering the random sample  according to the values of $X_i$s, denoted as $\{(X_{(i)}, Y_{(i)}), i = 1,\ldots,n\}$, where $X_{(1)}\leq \ldots\leq X_{(n)}$. 
The next step is dividing $\{(X_{(i)}, Y_{(i)}), i = 1,\ldots,n\}$ into $H$ slices, each with  $c$ observations. Here, $n = Hc$ is assumed for simplicity. 
Let the observations in the $h$-th slice be $\{(X_{(h, j)},  Y_{(h, j)}), j = 1,\ldots, c \}$. 
Denote $r_{(h, j)}$ as the number of $T_{i}$s such that $Y_{(h,j)}\geq T_i$, and $R_i$ as the number of $Y_j$s such that $Y_j\geq T_i$, for $i=1,\ldots, n$. The sliced estimates of  \eqref{definition:dependencemeasure} can be written as 
\beqr\label{definition:estimate}
\wh\calS(X,Y) &\defby &1 - (n - 1)(c-1)^{-1}\sum_{h=1}^{H}\sum_{j<l}^{c} \abs{r_{(h,j)}-r_{(h,l)}}\ \Bigg /\  \sum_{i = 1}^n R_i(n - R_i).
\eeqr
The sliced estimates reduce the computation complexity from $O(n^3)$ to $O\{n\log (n) \}$ compared to the previous kernel smoothing based estimates \citep{kong2019CompositeCoefficientDetermination}, and 
on the other hand, boost the power of the nearest neighbor estimates proposed in \cite{chatterjee2020NewCoefficientCorrelation}.
These estimates converge in distribution to standard normal distribution, when $X$ and $Y$ are independent \citep[Theorem 1 and 2]{zhang2021SlicedIndependenceTest}. 
We go further to show that the convergence is uniform as follows. 
Set $\Phi(\cdot)$ as the distribution function of standard normal distribution.


{\prop{\label{prop:normality}} When $X$ and $Y$ are independent, there exists $c_{0}>0$ such that,
\beqrs
\sup_{-\infty<t<\infty} \Big\lvert\pr\left[ \{n(c-1)\}^{1/2}\wh\calS(X,Y)/\sigma\leq t\right] - \Phi(t)\Big\rvert\leq c_{0}(c/n)^{1/3},
\eeqrs
for sufficiently large $n$. Here, $\theta_1 \defby E\Big[\cov^{2}\{1(Y_1\geq T), 1(Y_2\geq T)\mid T\}\Big]$, $\theta_2 \defby E\Big[\var\{1(Y\geq T)\mid T\}\Big]$ and $\sigma^2\defby 2 \theta_1/\theta_2^2$. 
}

When $Y$ is continuous and $T$ follows the same distribution as $Y$, we have $\sigma^2 = 4/5$. 
Compared to \cite{zhang2021SlicedIndependenceTest}, we establish the uniform convergence rate of the asymptotic normality to facilitate the FDR control in the screening procedure. The convergence rate of distribution is proportional to $(c/n)^{1/3}$. The smaller the number observations within slices $c$, the faster the convergence in distribution. 


\csubsection{Feature screening via sliced independence estimates}
As discussed in Section \ref{sec:sit}, the sliced independence estimates fulfill all implications,  model-free,  monotone invariant and computationally efficient. 
Therefore, we propose to employ the sliced independence estimates in the feature screening, which proceeds as follows. 
Define $\omega_k\defby\calS(X_k,Y)$  and $\wh\omega_k\defby\wh\calS(X_k,Y)$, for $k = 1,\ldots,p$, where $\wh\omega_k$ is  obtained when a random sample of size $n$, $\{ (\x_i, Y_i),i=1,\ldots, n\}$,  is available. We retain the covariates indexed by 
 \beqrs\label{equ:screening_threshold}
\wh\calA&\defby&\{k:\wh\omega_k\geq c_{1}n^{-\gamma}\},
\eeqrs
 for some $c_{1}>0$ and $0<\gamma<1/2$. 

We call the above screening procedure as \textit{SIT screening procedure}. The SIT screening procedure possesses the sure screening property under certain regularity conditions. Towards this goal, we introduce the following definitions illustrated in \cite{zhang2021SlicedIndependenceTest}. 
A family of real-valued functions, $x\mapsto f(t; x)$, for $t\in\calT$, has a uniform total variation of order $r$ over $\calT$, if for $B>0$, 
\beqr\label{definition:total_variation}
\lim_{n\to \infty} n^{-r}\sup_{t\in\calT, \Pi_n(B)}\sum_{i = 1}^n \abs{f(t; \wt X_{(i + 1)}) - f(t; \wt X_{(i)})} = 0, 
\eeqr
where  $\Pi_n(B)$ is a collection of all possible $n$-point partitions of $[-B, B]$ such that $-B\leq \wt X_{(1)}\leq \ldots \leq \wt X_{(n)}\leq B$.  
The uniform bounded variation condition is a special case for $r=0$. 
The function $x\mapsto f(t; x)$ is called non-expansive in the metric of $M(x)$ on both sides of $B_0$, if there exists a non-decreasing real-valued function $M(x)$ and a real number $B_0>0$, such that for any two points, say, $\wt X_1$ and $\wt X_2$, both in $(-\infty, -B_0]$ or both in $[B_0, \infty)$, 
\beqr\label{definition:non_expansive}
\abs{f(t; \wt X_1) - f(t; \wt X_2)} \leq \abs{M(\wt X_1) - M(\wt X_2)}.
\eeqr
Denote $s(t; X) \defby \pr (Y \ge t \mid X)$. We further provide the following conditions.
\begin{enumerate}[label=(C\arabic*$^{}$)]
	\item \label{condition:sis:total_variation} Assume that $x\mapsto s(t; x)$ has uniform probability bound on total variation of order $r>0$ and is non-expansive in the metric of $M(x)$ on both sides of a real number $B_0>0$ such that for any $x$,
	$x^b pr\{\abs{M(X)}>x\eta\} \leq \exp(-\eta^2)$ with some $b>0$.
	\item \label{condition:sis:slice_order}  Let $c = O(n^{\alpha})$, where $ 1-\max\{r,  1/b\} - \gamma > \alpha>0$.
	\item \label{condition:sis_order}  The minimal signal  of active covariates satisfies $\min\limits_{k\in\calA} \omega_{k}\geq 2c_{1}n^{-\gamma}$.
\end{enumerate}
Condition \ref{condition:sis:total_variation} bounds the tail probabilities. 
A sufficient condition for Condition \ref{condition:sis:slice_order} is that $x\mapsto s(t; x)$ is $L$-Lipschitz continuous and $X$ is sub-Gaussian or has a bounded support, in which case $c = o(n^{1-\gamma})$. Condition \ref{condition:sis_order} requires the minimal signal to be strong enough. Similar conditions are widely used in the screening literature. See, for example \cite{fan2008sure}, \cite{li2012feature}, \cite{kong2019CompositeCoefficientDetermination} and \cite{Liu2022}.
{\theo{\label{theorem:sis}} Suppose Conditions \ref{condition:sis:total_variation}--\ref{condition:sis:slice_order} hold.\\
\subtheo{\label{theorem:SIT_CR}}Convergence Rate. There exists  $c_{2}>0$ such that 
		\beqrs
		\pr(\max_{1\leq k \leq p}\abs{\wh\omega_{k} -\omega_{k}}\geq c_{1}n^{-\gamma})\leq  O\big\{p\exp(-c_{2}n^{\min\{1,2-2\max(1/b,r) - 2\alpha\}-2\gamma})\big\}.
		\eeqrs
  \subtheo{\label{theorem:sure_screening}} Sure Screening Property. 
Suppose Condition \ref{condition:sis_order} additionally holds. Denote $|\calA|$ as the cardinality of $\calA$. We have
		\beqr\label{equ:sure_screening}
		\pr(\calA \subseteq\wh\calA)\geq  1- O\big\{ |\calA| \exp(-c_{2}n^{\min\{1,2-2\max(1/b,r) - 2\alpha\}-2\gamma})\big\}.
		\eeqr
  \subtheo{\label{theorem:rank_consis}} Rank Consistency Property.
Suppose Condition \ref{condition:sis_order} additionally holds, and the signals of inactive covariates satisfy $\omega_k=0$ for any $k\in\calI$. We have
\beqrs
\pr(\min_{k\in\calA}\wh\omega_k-\max_{k\in\calI}\wh\omega_k>0)\geq 1-O\big\{p\exp(-c_{2}n^{\min\{1,2-2\max(1/b,r) - 2\alpha\}-2\gamma})\big\}.
\eeqrs
	}
Theorem~\ref{theorem:SIT_CR} states that $\wh\omega_k$ converges to $\omega_k$ exponentially with sample size $n$, which facilitates the remaining two properties in ultrahigh dimensional data analysis. 
Theorems~\ref{theorem:sure_screening} and \ref{theorem:rank_consis} indicate that all and only active covariates will be retained with probability approaching one when
\beqr\label{equ:p_n_perfect_thred}
p=o\big\{\exp(c_{2}n^{\min\{1,2-2\max(1/b,r) - 2\alpha\}-2\gamma})\big\},
\eeqr 
as sample size $n$  diverges to infinity. 
To be precise, besides \eqref{equ:sure_screening}, we also have
\beqrs
\pr(\calI \subseteq\wh\calA)\leq O\big\{ |\calI| \exp(-c_{2}n^{\min\{1,2-2\max(1/b,r) - 2\alpha\}-2\gamma})\big\}.
\eeqrs
This suggests that $c_1n^{-\gamma}$ is an optimal threshold, denoted as $L^\star$, because active and inactive covariates can be perfectly separated by it. 
However, $c_1$ and $\gamma$ is unattainable in practice, neither does the optimal threshold. 
An alternative of the optimal threshold, a data-adaptive threshold, is at demand.

\csection{Feature Screening with Data-adaptive Threshold}\label{sec:sisfdr}
\csubsection{SIT-BY Procedure}
In this section, we offer a data-adaptive threshold for the second step in screening procedure.
Given threshold $L\in\mathbb{R}$, we select active covariates with the index set 
\beqrs
\wh\calA(L)\defby\{k:\wh\omega_k\geq L, \textrm{ for } 1\leq k\leq p\}.
\eeqrs
A satisfactory threshold should be small enough to ensure that all active covariates included, that is 
$\calA\subseteq\wh\calA(L)$ holds with high probability.
Meanwhile, it can not be too small to restrict the inactive covariates involved. 
To  measure this degree of inactive covariates selected, we introduce \textit{false discovery 
rate} (FDR, \citealt{Benjamini1995}),and its empirical form, \textit{false discovery 
proportion} (FDP), which are defined as
\beqrs
\FDR(\wh\calA(L)) 
\defby
\mathbb{E}\left\{\FDP(\wh\calA(L)\right\}
\quad\text{and}\quad
\FDP(\wh\calA(L))
\defby
\frac{|\wh\calA(L) \cap \calI|}{\max(|\wh\calA(L)|, 1)},
\eeqrs 
respectively. The main purpose is to find the data-adaptive threshold $L$, so that FDR is controlled at a nominal level $q\in (0,1)$. 

Towards this goal, there has been numerous approaches, and they can be generally divided into two types. 
The first type constructs test statistics that are symmetric to zero for inactive covariates, but with sufficiently large values for active covariates. Benefit from the symmetric property, the number of false rejections 
can be estimated
\citep{Barber2015,Candes2018, Fan2020}
And such symmetric approach is often applied, when  the asymptotic distributions of marginal test statistics in screening are unknown \citep{Guo2021,Liu2022}. 
Another type of approaches in controlling FDR is based on 
the distribution of test statistics. 
Classical method is BH procedure, that can successfully control FDR when the test statistics satisfy \textit{positive regression dependence on a subset} (PRDS) condition \citep{Benjamini1995}.
To remove this dependence condition, the BY procedure, improves BH procedure with an adjusted nominal FDR level. Thus, it can control FDR no matter how test statistics depend on each other \citep{Benjamini2001}.
To further improve the detection power of distributional approaches, the dependency among test statistics should be fully or partially known
\citep{Fithian2020}.


In the first step of screening, sliced independence test stated in Proposition \ref{prop:normality} provides the essential asymptotic normality for marginal statistics. This largely urges us to adopt the distribution based approach in controlling FDR. As the dependency across $\{\wh\omega_k\}$ is nearly formidable to elucidate, BY procedure that deals with arbitrary dependency is high advocated.
Thus, we formally introduce the SIT-BY screening procedure, with data-adaptive threshold defined as
\beqr\label{equ:thrd_q}
L_q= \inf\left\{t>0:S(p)\frac{p\left\{1-\Phi\left(\{n(c-1)\}^{1/2}t/\sigma \right)\right\}}{|\{k:\wh\omega_k\geq t\}|\vee1}\leq q \right\},
\eeqr
where $S(p)=\sum_{l=1}^p 1/l$ is the BY-adjusted constant. Finally, the selected covariate set is $\wh\calA(L_q)$. The BY procedure is criticized to be too conservative, but this is not the case when the dependency among $\{\wh\omega_k\}$ is complicated and unknown. The necessity of the BY-adjusted constant will be demonstrated in Section~\ref{sec:simu_fdr} by comparing with setting $S(p)=1$ (SIT-BH procedure).

\csubsection{Sure Screening Property and Asymptotic FDR control}
We establish the theoretical properties of the SIT-BY procedure in terms of both  sure screening property and asymptotic FDR control in this subsection. To this end, we impose a mild condition about $(n,p)$.
\begin{enumerate}[label=(C\arabic*$^{}$)]
    \setcounter{enumi}{3}
	\item \label{condition:p_n_cond} $p\{1+\log(p)\} \exp(-c_1 n^{1-2\gamma}/\sigma^2)
\leq q|\calA|$.
\end{enumerate}
Condition \ref{condition:p_n_cond} generally holds. For example, when $\log(p)=o(  n^{1-2\gamma})$, the LHS of Condition \ref{condition:p_n_cond} tends to be zero and therefore tends to be less than $q|\calA|$ for any $q\in(0,1)$.
{\theo{\label{theorem:SIT_BY}} Consider SIT-BY procedure with a nominal FDR level $q\in(0,1)$. Under conditions \ref{condition:sis:total_variation}-\ref{condition:p_n_cond},\\
\subtheo{\label{theorem:SIT_BY:screening}} Sure Screening property. The SIT-BY procedure satisfies
\beqr
\pr(\calA\subseteq\wh\calA(L_q))\geq  1- O\big\{ |\calA| \exp(-c_{2}n^{\min\{1,2-2\max(1/b,r) - 2\alpha\}-2\gamma})\big\};
\eeqr
\subtheo{\label{theorem:SIT_BY:FDR}} Asymptotic FDR control. Letting $p\log(p)/|\calA|=o\{(n/c)^{1/3}\}$, we have
\beqrs
\lim_{(n,p)\rightarrow\infty} \FDR(\wh\calA(L_q))\leq q.
\eeqrs
}

Theorem~\ref{theorem:SIT_BY:screening} demonstrates all active covariates can be selected with probability tends to one with a reasonable $q$.
Therefore, in ultra-high dimensional, sure screening property can be established with our data-adaptive threshold. Theorem~\ref{theorem:SIT_BY:FDR} states our SIT-BY procedure can control FDR asymptotically, under moderate condition $p\log(p)/|\calA|=o\{(n/c)^{1/3}\}$. 
This condition requires the number of slices,  $H=n/c$, is sufficiently large to guarantee the converge rate of distribution. And special cases of this condition includes, (\rNum{1}) $n/c=o(p^3\{\log(p)\}^3)$; (\rNum{2}) the proportion of active covariates tends to be a constant strictly in (0,1), i.e., $|\calA|/p\rightarrow \pi_1\in(0,1)$ and $p=o[\exp\{(n/c)^{1/3}\}]$; or (\rNum{3}) $\lim\inf_{(n,p)\rightarrow}|\calA|/[p/\{\log(p)\}^{d_0}]\geq d_1>0$ and $p=o[\exp\{(n/c)^{1/3(1+d_0)}\}]$ for some $d_0, d_1>0$.   The last two cases are suitable for ultra-high dimension and fill the gap when the active covariates set is non-sparse. 
This condition is inevitable to ensure the convergence of distribution. And similar one was also imposed in \cite{Guo2021}, which suggested the necessity of a not too small $|\calA|$ in asymptotic FDR control. 
A noteworthy advantage of the SIT-BY procedure is it puts no restriction about the dependency among $\{\wh\omega_k\}$. 
Generally speaking, the SIT-BY procedure can satisfactorily recover the active covariates set.


\csection{Numerical Results}\label{sec:simu}
In this section, we investigate the finite sample performance of the SIT screening procedure through extensive simulations and an application of gene expression data.

\csubsection{SIT Screening Procedure}\label{sec:nr:sit}
We study the performance of the SIT screening procedure using the hard thresholding rule, mainly to show the effectiveness of using sliced independent test in screening procedure. 
In Study 1, we evaluate how the number of observations within each slice, $c$, affects the SIT screening performance. In Study 2, we compare the SIT screening procedure with four competing methods. 

\noindent\textbf{Study 1.}  
To examine the influence of different observations within each slice $c$, we compare the performance of the SIT screening procedures with $c$ varying in $\{2,4,8,16,32\}$.
Here, we generate $\x = (X_1,\ldots,X_p)\trans$ from multivariate normal distribution with mean zero and covariance matrix $\bSig =(\sigma_{kl})_{p\times p}$, where $\sigma_{kl} = \rho^{\abs{k-l}}$. 
We draw $\varepsilon_{\textrm{normal}}$ from standard normal distribution, and $\varepsilon_{\textrm{t}}$ from $t$ distribution with $3$ degrees of freedom. Then, we denote $\bb\defby(\vo_{s}, \vz_{p-s})\trans$, and consider the following two linear models and two generalized linear models.
\begin{enumerate}
	\item[\mylabel{model:a.1}{(a.1)}] $Y = \x\trans\bb + 2\varepsilon_{\textrm{normal}}$.
	\item[\mylabel{model:a.2}{(a.2)}] $Y = \x\trans\bb + 2\varepsilon_{\textrm{t}}$.
	\item[\mylabel{model:a.3}{(a.3)}] $Y = \exp(\x\trans\bb) + 4\varepsilon_{\textrm{normal}}$.
	\item[\mylabel{model:a.4}{(a.4)}] $Y = \exp(\x\trans\bb) + 4\varepsilon_{\textrm{t}}$.
 \end{enumerate}
We fix $n=256$, $p=1000$, $s = 4$ and $\rho = 0.5$. We repeat each experiment for $500$ times. 
Following \cite{li2012feature}, we evaluate the performance of a screening method with a default model size $d = \lfloor n/\{\log(n)\}\rfloor = 32$ through three criteria.
\begin{enumerate}
	\item[(a)] The proportion of selecting an individual active covariate for the fixed model size $d$ out of $500$ replications. 
	\item[(b)] $\calP_a$: The proportion of selecting all active covariates for the fixed model size $d$ out of $500$ replications.
	\item[(c)] The 25\%, 50\%, 75\% and 95\% quantiles of the minimum model sizes that select all active covariates out of $500$ replications.
\end{enumerate}

The simulation results are reported in Table \ref{table:sislinear}. 
A comparison of three criteria shows that a larger $c$ yields better performance in ranking the active covariates first, especially when  $\x$ and $Y$ are non-linear dependent and the distribution of error is heavily tailed. These results correspond to the Theorem 1 of \cite{zhang2021SlicedIndependenceTest}, which states that the larger $c$ is, the  more powerful the sliced independence test will be. The power improvement further helps us screen the active covariates.

\begin{table}[htp!]
		\scriptsize
		\captionsetup{font=footnotesize}
		\caption{
		Empirical results for Study 1: The proportions of selecting an individual active covariates or all covariates ($\calP_{a}$), and the 25\%, 50\%, 75\% and 95\% quantiles of the minimum model sizes out of $500$ replications. The screening methods include the SIT screening procedures with $c=2$ ($\textrm{SIT}_{2}$), $c=4$ ($\textrm{SIT}_{4}$), $c=8$ ($\textrm{SIT}_{8}$), $c=16$ ($\textrm{SIT}_{16}$) and $c=32$ ($\textrm{SIT}_{32}$).}
		\label{table:sislinear}    
		\begin{center}
			\begin{tabular}{l|rrrr|r|rrrr}
				\hline
				\hline
				&\multicolumn{9}{c}{Model (a.1)}\\
				\hline
				& $X_{1}$ & $X_{2}$ & $X_{20}$ & $X_{21}$ & $\calP_{a}$ &  25\% & 50\% & 75\% & 95\%  \\
				$\textrm{SIT}_{2}$  & 0.824 & 0.994 & 0.998 & 0.842 & 0.678 & 6 & 15 & 44 &166 \\
				$\textrm{SIT}_{4}$  & 0.988 & 1.000 & 1.000 & 0.990 & 0.978 & 4 &  4 &  5 & 12 \\
				$\textrm{SIT}_{8}$  & 1.000 & 1.000 & 1.000 & 1.000 & 1.000 & 4 &  4 &  4 &  4 \\
				$\textrm{SIT}_{16}$  & 1.000 & 1.000 & 1.000 & 1.000 & 1.000 & 4 &  4 &  4 &  4 \\
				$\textrm{SIT}_{32}$  & 1.000 & 1.000 & 1.000 & 1.000 & 1.000 & 4 &  4 &  4 &  4 \\
				\hline
			    &	\multicolumn{9}{c}{Model (a.2)}\\
			    \hline	
				& $X_{1}$ & $X_{2}$ & $X_{3}$ & $X_{20}$ & $\calP_{a}$ & 25\% & 50\% & 75\% & 95\%\\
				$\textrm{SIT}_{2}$  & 0.682 & 0.960 & 0.932 & 0.646 & 0.396 & 15 & 49 & 119 & 384\\
				$\textrm{SIT}_{4}$  & 0.950 & 1.000 & 0.998 & 0.954 & 0.904 & 4 & 5 & 11 & 62\\
				$\textrm{SIT}_{8}$  & 0.992 & 1.000 & 1.000 & 0.994 & 0.986 & 4 & 4 & 4 & 7\\
				$\textrm{SIT}_{16}$  & 0.998 & 1.000 & 1.000 & 1.000 & 0.998 & 4 & 4 & 4 & 4\\
				$\textrm{SIT}_{32}$  & 1.000 & 1.000 & 1.000 & 1.000 & 1.000 & 4 & 4 & 4 & 4\\
			     \hline
				&\multicolumn{9}{c}{Model (a.3)}\\
				\hline	
				& $X_{1}$ & $X_{2}$ & $X_{3}$ & $X_{20}$ & $\calP_{a}$ &  25\% & 50\% & 75\% & 95\%  \\
				$\textrm{SIT}_{2}$  & 0.638 & 0.956 & 0.948 & 0.624 & 0.370 & 21 & 50 & 144 & 357 \\
				$\textrm{SIT}_{4}$  & 0.918 & 1.000 & 1.000 & 0.946 & 0.866 & 4 & 6 & 13 & 93 \\
				$\textrm{SIT}_{8}$  & 0.996 & 1.000 & 1.000 & 0.992 & 0.988 & 4 & 4 & 4 & 9  \\
				$\textrm{SIT}_{16}$  & 1.000 & 1.000 & 1.000 & 0.996 & 0.996 & 4 & 4 & 4 & 4  \\
				$\textrm{SIT}_{32}$  & 1.000 & 1.000 & 1.000 & 1.000 & 1.000 & 4 & 4 & 4 & 4 \\
				\hline				
				& \multicolumn{9}{c}{Model (a.4)}\\
				\hline	
				& $X_{1}$ & $X_{2}$ & $X_{11}$ & $X_{12}$ & $\calP_{a}$ & 25\% & 50\% & 75\% & 95\% \\
				$\textrm{SIT}_{2}$  & 0.598 & 0.922 & 0.892 & 0.564 & 0.302 & 24 & 68 & 179 & 602 \\
				$\textrm{SIT}_{4}$  & 0.878 & 1.000 & 0.996 & 0.902 & 0.784 & 4 & 8 & 24 & 108\\
				$\textrm{SIT}_{8}$  & 0.982 & 1.000 & 1.000 & 0.992 & 0.974 & 4 & 4 & 5 & 16\\
				$\textrm{SIT}_{16}$  & 0.998 & 1.000 & 1.000 & 0.998 & 0.996 & 4 & 4 & 4 & 6\\
				$\textrm{SIT}_{32}$  & 1.000 & 1.000 & 1.000 & 1.000 & 1.000 & 4 & 4 & 4 & 4\\
				\hline
				\hline
			\end{tabular}
		\end{center}
\end{table}

\noindent\textbf{Study 2.} We compare the SIT screening procedures with $c=2$ ($\textrm{SIT}_{2}$) and $c=32$ ($\textrm{SIT}_{32}$) to the other popular screening methods: the distance correlation based screening (DC, \citealt{li2012feature}), the modified Blum-Kiefer-Rosenblatt correlation based screening (MBKR, \citealt{zhou2018ModelfreeFeatureScreening}),  Heller-Heller-Gorfine correlation  based screening (HHG, \citealt{heller2013ConsistentMultivariateTest}), the composite coefficient of determination based screening (CCD, \citealt{kong2019CompositeCoefficientDetermination}).
We follow the data generating strategy in Study 1, but generate $Y$ from four nonlinear models. The first three models contain interactions, heterogeneity, and exponential transformation. The last one contains two-way interaction terms. Similar models have been used in \cite{zhou2018ModelfreeFeatureScreening}.
\begin{enumerate}
	\item[\mylabel{model:b.1}{(b.1)}] $Y = 4X_{1}X_{2} + \exp\big\{5(X_{20}+X_{21})1(X_{20}+X_{21}\leq 3)\big\}\varepsilon_{\textrm{normal}}$.
	\item[\mylabel{model:b.2}{(b.2)}] $Y = 4X_{1}X_{2} + 3X_{3}^{2} + \exp\{5X_{20}1(X_{20}\leq 3)\}\varepsilon_{\textrm{normal}}$.
	\item[\mylabel{model:b.3}{(b.3)}] $Y = 4X_{1} + 5X_{2} + 3X_{3}^2 +   \exp\{5X_{20}1(X_{20}\leq 4)\}\varepsilon_{\textrm{normal}}$.
	\item[\mylabel{model:b.4}{(b.4)}] $Y = 2X_{1}X_{2} + 3X_{11}X_{12} + \varepsilon_{\textrm{normal}}$.
 \end{enumerate}

We fix $n=256$, $p=1000$, $\rho = 0.8$. We repeat each experiment for $500$ times. The default model size and three criteria are the same as those in Study 1. 
The simulation results are reported in Table \ref{table:sisnonlinear}. 
Generally speaking, the $\mathrm{SIT}_{32}$ performs the best in most settings, followed by the HHG, MBKR and CCD methods. For the $\mathrm{SIT}_{32}$, the medians and 95\% quantiles of minimum model sizes are four and seven, respectively, indicating over half of the replications correctly rank the four active covariates first, and meanwhile the 95\% replications rank them in the top seven. Other screening methods are non-comparable to the $\mathrm{SIT}_{32}$ in Model \ref{model:b.2}, in that other screening methods are 1.5 times as many minimum model sizes as $\mathrm{SIT}_{32}$ free of quantile levels. 

\begin{table}[htp!]
		\scriptsize
		\captionsetup{font=footnotesize}
		\caption{
		Empirical results for Study 2: 
		The proportions of selecting an individual active covariates or all covariates ($\calP_{a}$), and the 25\%, 50\%, 75\% and 95\% quantiles of the minimum model sizes out of $500$ replications. The screening methods include DC,  MBKR,   HHG,  CCD and the SIT screening procedures with $c=2$ ($\textrm{SIT}_{2}$) and $c=32$ ($\textrm{SIT}_{32}$).
		}
		\label{table:sisnonlinear}    
		\begin{center}
			\begin{tabular}{l|rrrr|r|rrrr}
				\hline
				\hline
				&\multicolumn{9}{c}{Model (b.1)}\\
				\hline
				& $X_{1}$ & $X_{2}$ & $X_{20}$ & $X_{21}$ & $\calP_{a}$ &  25\% & 50\% & 75\% & 95\%  \\
				DC    & 0.024  & 0.026 & 1.000 & 1.000  & 0.008  & 436 & 671 &  837 &  976  \\
				HHG   & 1.000  & 1.000 & 1.000 & 1.000  & 1.000 & 5   &  6  &  6  &  8  \\
				$\textrm{SIT}_{2}$  & 0.558  & 0.582 & 0.314 & 0.372  & 0.044    & 148 & 271 & 482 & 782 \\
				$\textrm{SIT}_{32}$ & 1.000  & 1.000 & 1.000 & 1.000  & 1.000   &  4  & 4   &  4  &  6  \\
				MBKR  & 1.000  & 1.000 & 1.000 & 1.000  & 1.000    &  6  & 7   &  10  &  17 \\
				CCD   & 1.000  & 1.000 & 1.000 & 1.000  & 1.000    &  5  &  6  &  6  &  8  \\ 
				\hline
			    &	\multicolumn{9}{c}{Model (b.2)}\\
			    \hline	
				& $X_{1}$ & $X_{2}$ & $X_{3}$ & $X_{20}$ & $\calP_{a}$ & 25\% & 50\% & 75\% & 95\%\\
				DC&  0.020  &  0.016  &  0.024  & 1.000 &  0.002   & 524   &   721 &  880  &  970  \\
             	HHG & 1.000  & 1.000 & 1.000 & 1.000  & 1.000   & 8   &  9  &  10  &  11  \\
				$\textrm{SIT}_{2}$ & 0.454  & 0.776 & 0.780 & 0.452  & 0.126    & 62 & 144 & 301  & 679  \\
			    $\textrm{SIT}_{32}$	 & 0.996  & 1.000 & 1.000 & 1.000  & 0.996    &  4  & 4   &  4  &  7  \\
				MBKR & 0.992  & 1.000 & 1.000 & 1.000  & 0.992    &  8  & 9   &  12  &  21 \\
				CCD & 1.000  & 1.000 & 1.000 & 1.000  & 1.000     &  7  &  8  &  10  &  12  \\
			     \hline
				&\multicolumn{9}{c}{Model (b.3)}\\
				\hline	
				& $X_{1}$ & $X_{2}$ & $X_{3}$ & $X_{20}$ & $\calP_{a}$ &  25\% & 50\% & 75\% & 95\%  \\
				DC    & 0.032  & 0.046 & 0.032 & 1.000  & 0.004   & 409 & 623 &  796 &  942  \\
				HHG   & 1.000  & 1.000 & 1.000 & 1.000  & 1.000   & 6   &  6  &  8  &  9  \\
				$\textrm{SIT}_{2}$  & 0.940  & 0.978 & 0.694 & 0.396  & 0.236   & 35 & 102 & 250 & 636 \\
				$\textrm{SIT}_{32}$ & 1.000  & 1.000 & 1.000 & 1.000  & 1.000   &  4  & 4   &  5  &  5\\
				MBKR  & 1.000  & 1.000 & 1.000 & 1.000  & 1.000   &  4  & 4   &  4  &  6\\
				CCD   & 1.000  & 1.000 & 1.000 & 1.000  & 1.000   &  4  &  5  &  6  &  7\\
				\hline				
				& \multicolumn{9}{c}{Model (b.4)}\\
				\hline	
				& $X_{1}$ & $X_{2}$ & $X_{11}$ & $X_{12}$ & $\calP_{a}$ & 25\% & 50\% & 75\% & 95\% \\
				DC &  1.000 & 1.000 & 1.000 & 1.000  & 1.000  & 4   &  4  &  6  &  9 \\
				HHG & 1.000  & 1.000 & 1.000 & 1.000  & 1.000  & 4   &  4  &  5  &  6\\
				$\textrm{SIT}_{2}$ & 0.782  & 0.804 & 0.996 & 0.988  & 0.628  & 8   &  20  & 66  &  218 \\
				$\textrm{SIT}_{32}$ & 1.000  & 1.000 & 1.000 & 1.000  & 1.000  &  4  & 4   &  4  &  6 \\
				MBKR & 1.000  & 1.000 & 1.000 & 1.000  & 1.000  & 4   &  4  &  5  &  6   \\
				CCD & 1.000  & 1.000 & 1.000 & 1.000  & 1.000  & 4   &  4  &  5  &  6   \\
				\hline
				\hline
			\end{tabular}
		\end{center}
\end{table}

\csubsection{FDR Control Performance}\label{sec:simu_fdr}
In this section, we demonstrate the effectiveness of our data-adaptive threshold from the performance of both the FDR control and sure screening property. 

\noindent\textbf{Study 3.} 
In this study, we assess three data-adaptive thresholds. These thresholds are based on the Benjamini-Yekutieli procedure (BY, \citealt{Benjamini2001}), the Benjamini-Hochberg procedure (BH, \citealt{Benjamini1995}) and the threshold sample splitting strategy with splitting parameter $K=4$ (SS, \citealt{Guo2021}). We fix $c=32$ for the SIT screening procedure.
We follow the data generating strategy in Study 1, but generate $Y$ from the following two linear models and two generalized linear model.
\begin{enumerate}
	\item[\mylabel{model:c.1}{(c.1)}] $Y =  2\x\trans\bb + \varepsilon_{\textrm{normal}}$.
	\item[\mylabel{model:c.2}{(c.2)}] $Y =  2\x\trans\bb + \varepsilon_{\textrm{t}}$.
	\item[\mylabel{model:c.3}{(c.3)}] $Y = \exp(\x\trans\bb/5) + \varepsilon_{\textrm{normal}}$.
	\item[\mylabel{model:c.4}{(c.4)}] $Y = \exp(\x\trans\bb/5) + \varepsilon_{\textrm{t}}$.
 \end{enumerate}
We fix $n=1024$, $p=5000$, $s=20$, $\rho = 0.5$. We repeat each experiment for 100 times with nominal FDR levels varying in $\{0.1,0.2\}$. 
Similar to Studies 1 and 2, we report the proportions of selecting an individual covariate out of $100$ replications.
Following \cite{Guo2021}, we further evaluate the data-adaptive thresholds regarding the FDR control and screening through two additional criteria.
\begin{enumerate}
	\item[(d)] AMS: Average model size induced by the data-adaptive threshold out of $100$ replications.
	\item[(e)] FDP: Average false discovery proportion induced by the data-adaptive threshold out of $100$ replications.
\end{enumerate}
Tables \ref{table:study3_1} and \ref{table:study3_2} report the results of the linear models \ref{model:c.1} and \ref{model:c.2} and the generalized linear models \ref{model:c.3} and \ref{model:c.4}, respectively. 
The SIT-BY procedure generally achieves the best FDR and screening trade-off, as it can select most of the 20 active covariates and control the FDP around the nominal FDR level.
The SIT-BH procedure faces severe FDR inflation because the BH procedure fails in handling the complex dependence among the marginal utilities. This FDR inflation sustains the necessity of the BY-adjusted constant. 
As for the SIT-SS procedure, though it always provides reliable FDR control, the screening performance is undesirable. A possible reason is the information loss caused by ignoring the asymptotic distribution and sample splitting. 
It is also worth mentioning that the average model sizes induced by the SIT-BY procedure are just slightly larger than $20$, indicating our method can satisfactorily recover the true model. By contrast, the STI-BH procedure misspecifies the model resulting from including too many irrelevant covariates. On the other hand, the SIT-SS procedure can only select about half of the covariates in Models \ref{model:c.3} and \ref{model:c.4}. Overall, all procedures perform well in the linear models, possibly because the dependency between covariates and response is easier to be captured.


\begin{table}[htp!]
		\scriptsize
		\captionsetup{font=footnotesize}
		\caption{Empirical results in Models \ref{model:c.1} and \ref{model:c.2}: The proportions of selecting an individual active covariates, average model size, and average false discovery proportion. The FDR level $q$ varies in $\{0.1, 0.2\}$. The data-adaptive thresholds are based on the Benjamini-Yekutieli procedure (BY, \citealt{Benjamini2001}), the Benjamini-Hochberg procedure (BH, \citealt{Benjamini1995}) and the sample splitting strategy with splitting parameter $K=4$ (SS, \citealt{Guo2021}).}
		\label{table:study3_1}
		\begin{center}
     \begin{tabular}{cc|llllllllll|l}
   \hline
   \hline
    \multicolumn{13}{c}{Model (c.1)}\\
   \hline
   \multirow{8}{*}{0.1} &       & $X_1$  & $X_2$    & $X_3$    & $X_4$    & $X_5$    & $X_6$    & $X_7$    & $X_8$    & $X_9$    & $X_{10}$   & FDP\\
    & BY    & 0.99 & 1.00 & 1.00 & 1.00 & 1.00 & 1.00 & 1.00 & 1.00 & 1.00 & 1.00 & 0.13\\
    & BH    & 0.99 & 1.00 & 1.00 & 1.00 & 1.00 & 1.00 & 1.00 & 1.00 & 1.00 & 1.00 & 0.41\\
    & SS & 0.35 &  0.58 &  0.74 &  0.74 &  0.75 &  0.77 &  0.81 &  0.80 &  0.76 &  0.74 &  0.01\\
    \cline{2-13} 
    &       & $X_{11}$  & $X_{12}$  & $X_{13}$  & $X_{14}$  & $X_{15}$  & $X_{16}$  & $X_{17}$  & $X_{18}$  & $X_{19}$  & $X_{20}$ & AMS \\
    & BY    & 1.00 & 1.00 & 1.00 & 1.00 & 1.00 & 1.00 & 1.00 & 1.00 & 1.00 & 0.99 & 23.22\\
    & BH    & 1.00 & 1.00 & 1.00 & 1.00 & 1.00 & 1.00 & 1.00 & 1.00 & 1.00 & 0.99 & 34.52\\
    & SS & 0.75 &  0.78 &  0.79 &  0.74 &  0.73 &  0.76 &  0.74 &  0.68 &  0.64 &  0.42 &  14.34\\
    \hline
    \multirow{8}{*}{0.2} &        & $X_1$  & $X_2$    & $X_3$    & $X_4$    & $X_5$    & $X_6$    & $X_7$    & $X_8$    & $X_9$    & $X_{10}$ & FDP \\
    & BY    & 0.99 & 1.00 & 1.00 & 1.00 & 1.00 & 1.00 & 1.00 & 1.00 & 1.00 & 1.00 & 0.20\\
    & BH    & 0.99 & 1.00 & 1.00 & 1.00 & 1.00 & 1.00 & 1.00 & 1.00 & 1.00 & 1.00 & 0.58 \\
    & SS & 0.71 &  0.92 &  0.94 &  0.94 &  0.95 &  0.94 &  0.99 &  0.98 &  0.96 &  0.99 &  0.05\\
    \cline{2-13} 
    &       & $X_{11}$  & $X_{12}$  & $X_{13}$  & $X_{14}$  & $X_{15}$  & $X_{16}$  & $X_{17}$  & $X_{18}$  & $X_{19}$  & $X_{20}$ & AMS \\
    & BY    & 1.00 & 1.00 & 1.00 & 1.00 & 1.00 & 1.00 & 1.00 & 1.00 & 1.00 & 0.99 & 25.25\\
    & BH    & 1.00 & 1.00 & 1.00 & 1.00 & 1.00 & 1.00 & 1.00 & 1.00 & 1.00 & 0.99 & 48.27\\
    & SS & 0.94 &  0.96 &  0.92 &  0.94 &  0.90 &  0.95 &  0.92 &  0.92 &  0.94 &  0.75 &  19.58\\
    \hline
        \multicolumn{13}{c}{Model (c.2)}\\
   \hline
    \multirow{8}{*}{0.1} &       & $X_1$  & $X_2$    & $X_3$    & $X_4$    & $X_5$    & $X_6$    & $X_7$    & $X_8$    & $X_9$    & $X_{10}$  & FDP \\
    & BY   & 0.99 &  1.00 &  1.00 &  1.00 &  1.00 &  1.00 &  1.00 &  1.00 &  1.00 &  1.00 & 0.13 \\
    & BH    &  1.00 &  1.00 &  1.00 &  1.00 &  1.00 &  1.00 &  1.00 &  1.00 &  1.00 &  1.00 & 0.41\\
    & SS    &  0.40 &  0.59 & 0.73 &  0.75 &  0.78 &  0.78 &  0.81 &  0.80 &  0.77 &  0.76 &  0.01\\
    \cline{2-13} 
    &       & $X_{11}$  & $X_{12}$  & $X_{13}$  & $X_{14}$  & $X_{15}$  & $X_{16}$  & $X_{17}$  & $X_{18}$  & $X_{19}$  & $X_{20}$ & AMS \\
    & BY    & 1.00 &  1.00 &  1.00 &  1.00 &  1.00 &  1.00 &  1.00 &  1.00 &  1.00 &  0.98 & 23.17 \\
    & BH     & 1.00 &  1.00 &  1.00 &  1.00 &  1.00 &  1.00 &  1.00 &  1.00 &  1.00 &  0.99 & 34.24\\
    & SS    & 0.76 &  0.78 & 0.79 &  0.77 &  0.74 &  0.78 &  0.76 &  0.72 &  0.67 &  0.40 &  14.60\\
    \hline
    \multirow{8}{*}{0.2} &        & $X_1$  & $X_2$    & $X_3$    & $X_4$ & $X_5$    & $X_6$    & $X_7$    & $X_8$    & $X_9$    & $X_{10}$ & FDP \\
    & BY    & 0.99 &  1.00 &  1.00 &  1.00 &  1.00 &  1.00 &  1.00 &  1.00 &  1.00 &  1.00 & 0.19\\
    & BH     & 0.99 &  1.00 &  1.00 &  1.00 &  1.00 &  1.00 &  1.00 &  1.00 &  1.00 &  1.00 & 0.57 \\  
    & SS    & 0.73 &  0.91 &  0.92 &  0.95 &  0.94 &  0.95 &  0.99 &  0.98 &  0.96 &  0.99 &  0.05\\
    \cline{2-13} 
    &       & $X_{11}$  & $X_{12}$  & $X_{13}$  & $X_{14}$  & $X_{15}$  & $X_{16}$  & $X_{17}$  & $X_{18}$  & $X_{19}$  & $X_{20}$ & AMS \\
    & BY    & 1.00 &  1.00 &  1.00 &  1.00 &  1.00 &  1.00 &  1.00 &  1.00 &  1.00 &  0.99 & 24.97\\
    & BH     & 1.00 &  1.00 &  1.00 &  1.00 &  1.00 &  1.00 &  1.00 &  1.00 &  1.00 &  0.99 & 47.76\\
    & SS    & 0.95 &  0.96 &  0.93 &  0.95 &  0.93 &  0.97 &  0.91 &  0.93 &  0.94 &  0.76 &  19.75\\
    \hline
    \hline
    \end{tabular}
    \end{center}
\end{table}

\begin{table}[htp!]
		\scriptsize
		\captionsetup{font=footnotesize}
		\caption{Empirical results in Models \ref{model:c.3} and \ref{model:c.4}: The proportions of selecting an individual active covariates, average model size, and average false discovery proportion. The FDR level $q$ varies in $\{0.1, 0.2\}$. The data-adaptive thresholds are based on the Benjamini-Yekutieli procedure (BY, \citealt{Benjamini2001}), the Benjamini-Hochberg procedure (BH, \citealt{Benjamini1995}) and the sample splitting strategy with splitting parameter $K=4$ (SS, \citealt{Guo2021}).}
		\label{table:study3_2}
		\begin{center}
     \begin{tabular}{cc|llllllllll|l}
   \hline
   \hline
    \multicolumn{13}{c}{Model (c.3)}\\
   \hline
   \multirow{8}{*}{0.1} &       & $X_1$  & $X_2$    & $X_3$    & $X_4$    & $X_5$    & $X_6$    & $X_7$    & $X_8$    & $X_9$    & $X_{10}$   & FDP\\
    & BY    & 0.72 &  0.97 &  1.00 &  1.00 &  1.00 &  1.00 &  1.00 &  1.00 &  1.00 &  1.00 &  0.13\\
    & BH    & 0.82 &  1.00 &  1.00 &  1.00 &  1.00 &  1.00 &  1.00 &  1.00 &  1.00 &  1.00 &  0.42\\
    & SS & 0.11 &  0.31 &  0.40 &  0.42 &  0.47 &  0.47 &  0.48 &  0.47 &  0.47 &  0.46 &  0.01\\
    \cline{2-13} 
    &       & $X_{11}$  & $X_{12}$  & $X_{13}$  & $X_{14}$  & $X_{15}$  & $X_{16}$  & $X_{17}$  & $X_{18}$  & $X_{19}$  & $X_{20}$ & AMS \\
    & BY    & 1.00 &  1.00 &  1.00 &  1.00 &  1.00 &  1.00 &  1.00 &  1.00 &  0.99 &  0.74 &  22.52\\
    & BH    & 1.00 &  1.00 &  1.00 &  1.00 &  1.00 &  1.00 &  1.00 &  1.00 &  1.00 &  0.84 &  34.93\\
    & SS &  0.48 &  0.44 &  0.47 &  0.46 &  0.46 &  0.47 &  0.44 &  0.38 &  0.28 &  0.12 &  8.24\\
    \hline
    \multirow{8}{*}{0.2} &        & $X_1$  & $X_2$    & $X_3$    & $X_4$    & $X_5$    & $X_6$    & $X_7$    & $X_8$    & $X_9$    & $X_{10}$ & FDP \\
    & BY    & 0.74 &  0.99 &  1.00 &  1.00 &  1.00 &  1.00 &  1.00 &  1.00 &  1.00 &  1.00 &  0.18 \\
    & BH    & 0.85 &  1.00 &  1.00 &  1.00 &  1.00 &  1.00 &  1.00 &  1.00 &  1.00 &  1.00 &  0.59 \\
     & SS & 0.26 &  0.56 &  0.79 &  0.81 &  0.86 &  0.85 &  0.87 &  0.84 &  0.85 &  0.85 &  0.03 \\
    \cline{2-13} 
    &       & $X_{11}$  & $X_{12}$  & $X_{13}$  & $X_{14}$  & $X_{15}$  & $X_{16}$  & $X_{17}$  & $X_{18}$  & $X_{19}$  & $X_{20}$ & AMS \\
    & BY    & 1.00 &  1.00 &  1.00 &  1.00 &  1.00 &  1.00 &  1.00 &  1.00 &  0.99 &  0.78 &  23.95\\
    & BH    & 1.00 &  1.00 &  1.00 &  1.00 &  1.00 &  1.00 &  1.00 &  1.00 &  1.00 &  0.86 &  48.99\\
    & SS & 0.80 &  0.79 &  0.81 &  0.83 &  0.85 &  0.82 &  0.77 &  0.72 &  0.67 &  0.28 &  15.43\\
    \hline
        \multicolumn{13}{c}{Model (c.4)}\\
   \hline
    \multirow{8}{*}{0.1} &       & $X_1$  & $X_2$    & $X_3$    & $X_4$    & $X_5$    & $X_6$    & $X_7$    & $X_8$    & $X_9$    & $X_{10}$  & FDP \\
    & BY   &  0.58 &  0.89 &  1.00 &  1.00 &  1.00 &  1.00 &  1.00 &  1.00 &  1.00 &  0.99 &  0.12 \\
    & BH   &  0.77 &  0.95 &  1.00 &  1.00 &  1.00 &  1.00 &  1.00 &  1.00 &  1.00 &  1.00 &  0.41 \\
    & SS &  0.08 &  0.18 &  0.33 &  0.38 &  0.40 &  0.44 &  0.46 &  0.44 &  0.46 &  0.43 &  0.01  \\
    \cline{2-13} 
    &       & $X_{11}$  & $X_{12}$  & $X_{13}$  & $X_{14}$  & $X_{15}$  & $X_{16}$  & $X_{17}$  & $X_{18}$  & $X_{19}$  & $X_{20}$ & AMS \\
    & BY    &  1.00 &  1.00 &  1.00 &  1.00 &  1.00 &  1.00 &  1.00 &  1.00 &  0.91 &  0.57 &  21.78 \\
    & BH   & 1.00 &  1.00 &  1.00 &  1.00 &  1.00 &  1.00 &  1.00 &  1.00 &  0.95 &  0.75 &  33.70 \\
    & SS &   0.40 &  0.46 &  0.44 &  0.44 &  0.41 &  0.42 &  0.36 &  0.37 &  0.25 &  0.12 &  7.41 \\
    \hline
    \multirow{8}{*}{0.2} &        & $X_1$  & $X_2$    & $X_3$    & $X_4$ & $X_5$    & $X_6$    & $X_7$    & $X_8$    & $X_9$    & $X_{10}$ & FDP \\
     & BY    & 0.64 &  0.92 &  1.00 &  1.00 &  1.00 &  1.00 &  1.00 &  1.00 &  1.00 &  0.99 &  0.18 \\
    & BH     & 0.81 &  0.97 &  1.00 &  1.00 &  1.00 &  1.00 &  1.00 &  1.00 &  1.00 &  1.00 &  0.58 \\
    & SS    & 0.21 &  0.43 &  0.66 &  0.73 &  0.72 &  0.75 &  0.77 &  0.74 &  0.77 &  0.74 &  0.04 \\
    \cline{2-13} 
    &       & $X_{11}$  & $X_{12}$  & $X_{13}$  & $X_{14}$  & $X_{15}$  & $X_{16}$  & $X_{17}$  & $X_{18}$  & $X_{19}$  & $X_{20}$ & AMS \\
    & BY    & 1.00 &  1.00 &  1.00 &  1.00 &  1.00 &  1.00 &  1.00 &  1.00 &  0.93 &  0.59 &  23.61 \\
    & BH    & 1.00 &  1.00 &  1.00 &  1.00 &  1.00 &  1.00 &  1.00 &  1.00 &  0.98 &  0.81 &  47.79 \\
    & SS    & 0.72 &  0.74 &  0.72 &  0.74 &  0.70 &  0.79 &  0.72 &  0.63 &  0.56 &  0.21 &  13.85 \\
    \hline
    \hline
    \end{tabular}
    \end{center}
\end{table}

\csubsection{Real data analysis}
We evaluate our proposed method through rat eye gene expression data to develop scientific insights into human eye disease \citep{scheetz2006regulation}. The dataset is available at \url{http://www.ncbi.nlm.nih.gov/geo} with Series ID GSE5680.
This dataset consists of 120 rat microarrays, each with 18,976 gene probes that exhibited sufficient signal for reliable analysis and at least 2-fold variation in expression level among the 120 rats. 
\cite{chiang2006homozygosity} found that Gene TRIM32 is highly correlated to Bardet–Biedl syndrome, a retinal disease, by analyzing SNP microarray genotyping of a small consanguineous family.
One of 18,976 rat probes, 1389163\_at, is from TRIM32, which means we can explore human eye disease probes by studying rat eye gene expression.
Our goal now turns to identify the probes that display similar patterns to the probe 1389163\_at among the remaining 18,975 probes. 

\begin{figure}[htbp]
		\captionsetup{font = footnotesize}
		\centering
		\subfigure[Original data]{\label{fig:realdata:a}
			\begin{minipage}[t]{0.45\linewidth}
				\centering
	 	\includegraphics[width=2.5in]{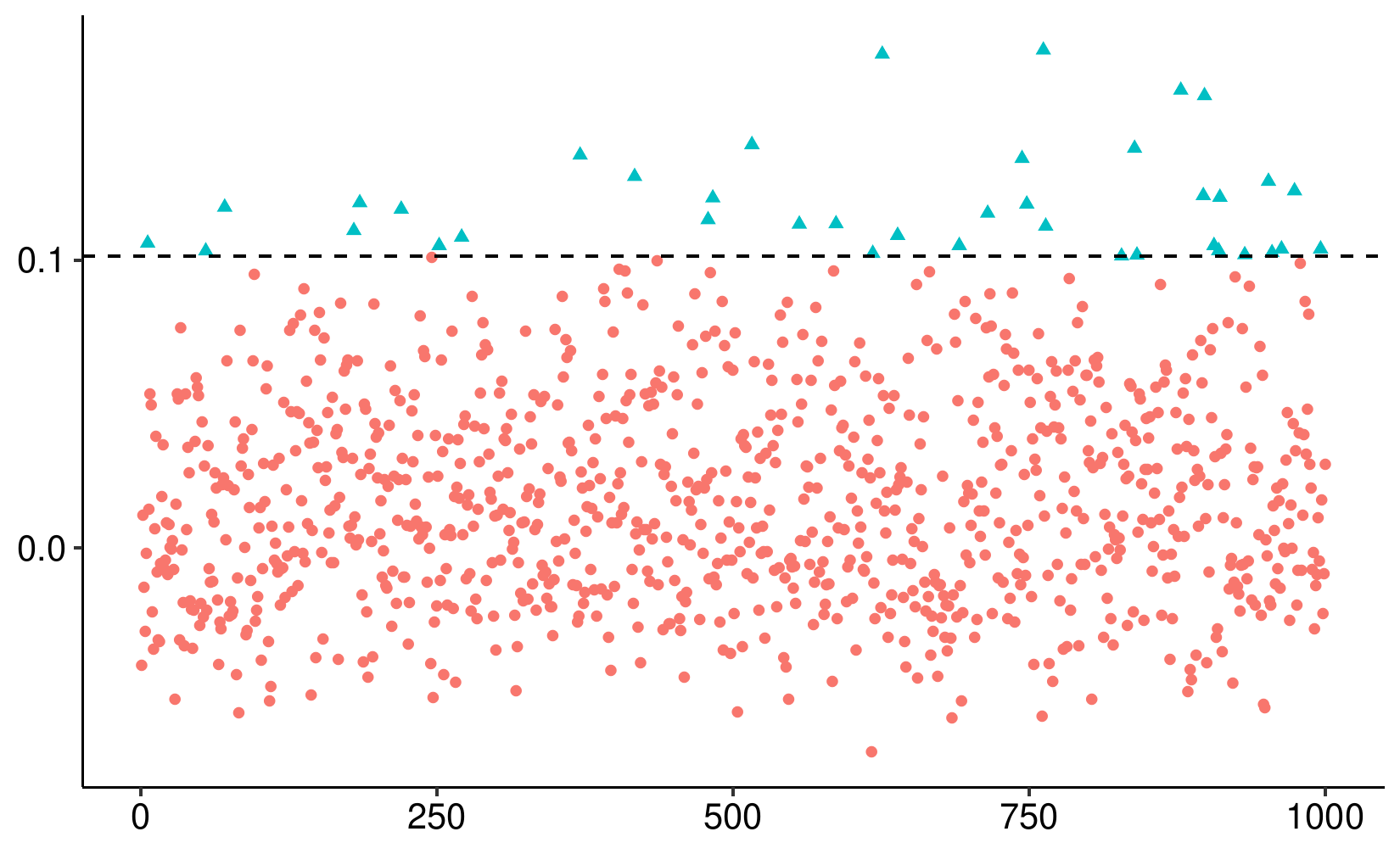}
			\end{minipage}%
		}%
		\subfigure[Original and auxiliary data]{\label{fig:realdata:b}
			\begin{minipage}[t]{0.45\linewidth}
				\centering
	       	\includegraphics[width=2.5in]{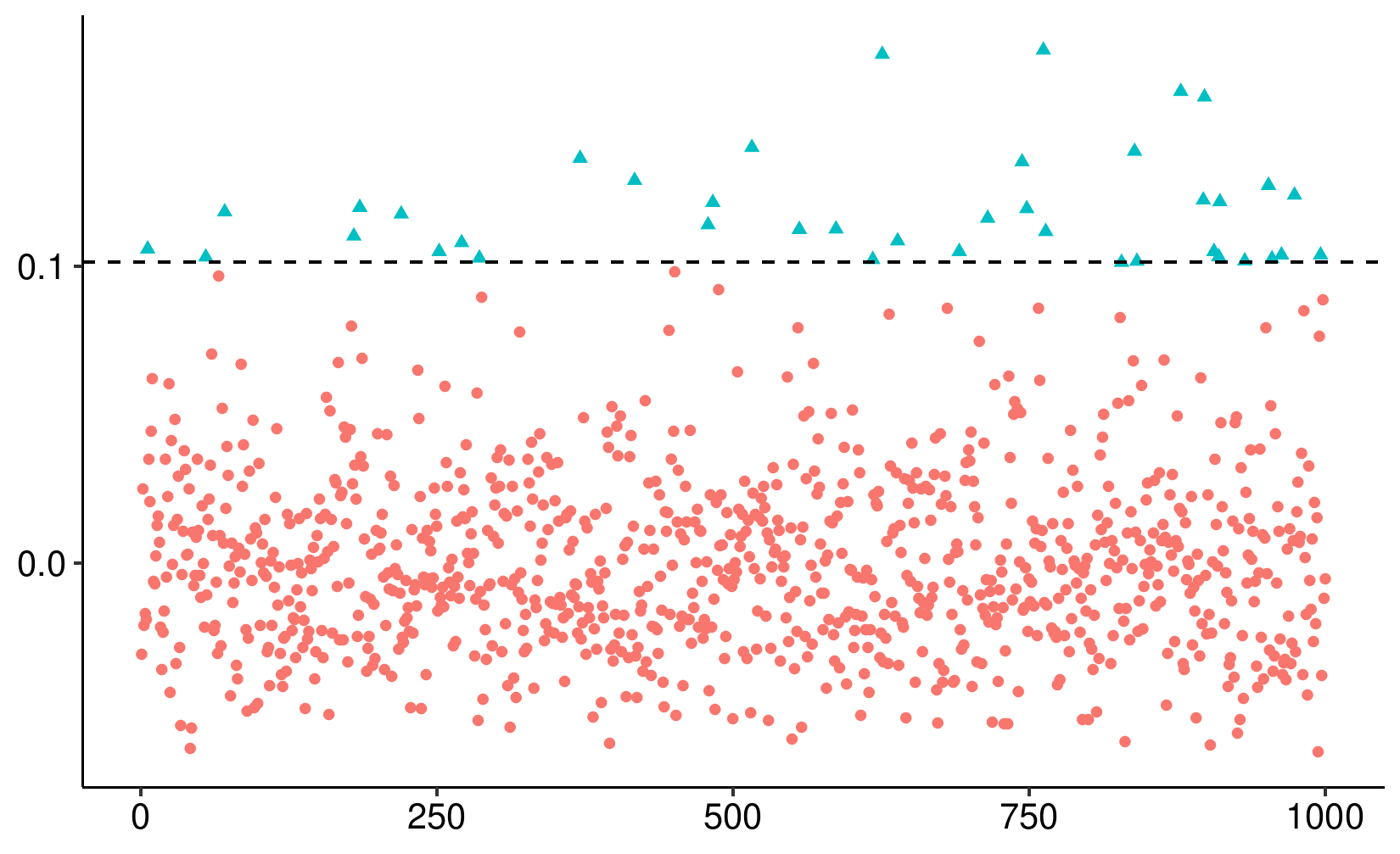}
			\end{minipage}%
		}%
		\centering
		\caption{Scatterplots of $\wh\omega_k$ obtained from sliced independence test with $c=8$ ($\mathrm{SIT}_{8}$).  
		In both subfigures, the horizontal axis represents 1000 probes and the vertical axis represents $\wh\omega_k$. 
		The horizontal dashed lines are the data-adaptive thresholds, above which are the selected probes ($\triangle$) and below which are unselected probes ($\circ$).
		}
		\label{fig:realdata}
	\end{figure}

Following \cite{zhong2015iterative}, we first select 1000 probes with the largest variances from the 18,975 probes. Then, we standardize these 1000 probes, so that each has zero mean and unit variance.
We apply our data-adaptive screening procedure, the SIT-BY procedure, with the nominal FDR level $q=0.1$, and identify 39 probes.
The 1000 sliced independence correlations and the data-adaptive threshold are shown in Figure \ref{fig:realdata:a}.
We can see that the selected probes, marked as triangle, tend to have larger correlations compared to the unselected ones, marked using circle. 
 
To further justify our selected probes, we replace those unselected probes with independent variables from standard normal distribution. 
We conduct the SIT-BY procedure on the combination of original and auxiliary data with the same nominal FDR level $q=0.1$. 
In this time, we select 40 probes, including all 39 probes previously chosen and one artificial probe.
Comparing Figures~\ref{fig:realdata:a} and \ref{fig:realdata:b}, we can find that the threshold remains almost unchanged. 
The stable threshold indicates that the selection rule, determined by original data, tends to include all dependent probes while permitting just a small number of independent ones. 
In contrast, the traditional hard thresholding rule selects $\lfloor n/\log(n) \rfloor = 25$ probes and risks omitting important probes.

\csection{Concluding Remarks}\label{sec:conclusion}
Research into feature screening under ultrahigh-dimensional setup has grown over the last two decades. Here we propose a sliced independence test based feature screening procedure and provide a general strategy for determining a data-adaptive threshold. 
The proposed SIT-BY procedure is model-free, monotone invariant and computational efficient. With our suggested data-adaptive threshold, it enjoys sure screening and rank consistency properties under mild conditions. In addition, the SIT-BY procedure asymptotically controls the FDR under a prespecified level $q\in(0,1)$ under a slight stringent but necessary condition. Simulation studies suggest that our proposed procedure is effective in ranking covariates and delivers the best FDR and screening trade-off. 

Conventional feature screening procedures establish the theoretical guarantee of sure screening property with an unknown threshold. A hard thresholding rule, however, is developed for practical consideration. Our proposed data-adaptive threshold fills the gap between theory and practice, as we further prove sure screening property for the SIT screening procedure with it. Moreover, the SIT-BY procedure avoids including too many redundant covariates. Generally speaking, our proposed method satisfactorily recovers the true active covariates set. 
\begin{center}
	\bibliography{reference}
\end{center}


\setcounter{section}{0} 
\renewcommand{\theequation}{\Alph{section}.\arabic{equation}} 
\renewcommand{\thetable}{\Alph{section}.\arabic{table}} 
\renewcommand{\thesection}{\Alph{section}}
\renewcommand{\thetheo}{\Alph{section}.\arabic{theo}}
\renewcommand{\theprop}{\Alph{section}.\arabic{prop}}
\renewcommand{\thecoll}{\Alph{section}.\arabic{coll}}
\renewcommand{\thelemm}{\Alph{section}.\arabic{lemm}}


\newpage
\begin{center}
	{\bf\large  Supplement to ``Model-Free, Monotone Invariant and Computationally Efficient Feature Screening with Data-adaptive Threshold"}
\end{center}

In this Supplement Material, we provide proofs for Theorems \ref{theorem:sis}-\ref{theorem:SIT_BY} and Proposition \ref{prop:normality} in the main context.  
For notation clarity, we define
\beqrs
\sum_{(j,l)}^{c}&\defby& \sum_{j=1}^{c} \sum_{l\neq j}^{c}.
\eeqrs 
We define $s_i(X_{(h,j)}) \defby E \{1(Y_{(h,j)} \ge T_{i}) \mid X_{(h,j)},T_{i}\}$, $\varepsilon_i(X_{(h,j)}) \defby 1(Y_(h,j)\geq T_i) - s_i(X_{(h,j)})$ and $\nu_{i}(Y_{j})\defby 1(Y_j\geq T_i) - E\{1(Y_j\geq T_i)\mid T_i\}$.
For simplicity, we further denote $W\defby E\Big[\var\{1(Y\geq T)\mid T\}\Big]$ and $\Lambda \defby E\Big[\var\{1(Y\geq T)\mid X,T\}\Big]$. Let $E_{T}$ be the expectation taking over the subscript $T$ given all other random variables.

\section{Proof of Theorem \ref{theorem:sis}}
\label{sec:theorem:sis}
In the following proof, we build concentration inequality for $\wh\calS(X, Y)$ and establish sure screening property based on this inequality.
We define
\beqrs
\wh\Lambda&\defby& \{n^2(c-1)\}^{-1}\sum\limits_{i=1}^{n}\sum_{h=1}^{H}\sum_{j<l}^{c}\{1(Y_{(h,j)}\geq T_i)-1(Y_{(h,l)}\geq T_i)\}^2\textrm{ and }\\
\wh{W}&\defby&\{n^2(n-1)\}^{-1}\sum_{i=1}^{n}\sum_{j<l}^{n}\{1(Y_{j}\geq T_i) - 1(Y_{l}\geq T_i)\}^2.
\eeqrs 
It follows that $\calS(X,Y) = (W-\Lambda)/W$ and $\wh\calS(X,Y) = (\wh{W}-\wh\Lambda)/\wh{W}$. Therefore,
\beqrs
n^{1/2}\{\wh\calS(X,Y) - \calS(X,Y)\} 
&=& n^{1/2}\{(\wh{W}-W)\Lambda - W(\wh{\Lambda} - \Lambda)\}/(\wh{W}W).
\eeqrs

Following the proof in \citet[Theorem 1]{zhang2021SlicedIndependenceTest}, we divide $\wh\Lambda$ into three parts. Define
\beqrs
I_{1} &\defby& \{n^2(c-1)\}^{-1}\sum_{i=1}^{n}\sum_{h=1}^{H}\sum_{j<l}^{c} \{s_i(X_{(h,j)}) - s_i(X_{(h,l)})\}^2,\\
I_{2} &\defby& \{n^2(c-1)\}^{-1}\sum_{i=1}^{n}\sum_{h=1}^{H}\sum_{j<l}^{c}\{s_i(X_{(h,j)}) - s_i(X_{(h,l)})\}\{\varepsilon_i(X_{(h,j)}) - \varepsilon_i(X_{(h,l)})\}\textrm{ and }\\
I_{3} &\defby& \{n^2(c-1)\}^{-1}\sum_{i=1}^{n}\sum_{h=1}^{H}\sum_{j<l}^{c}\{\varepsilon_i(X_{(h,j)}) - \varepsilon_i(X_{(h,l)})\}^2. 
\eeqrs
They further divide $I_{3} = I_{4} - I_{5}$, where
\beqrs
I_4&\defby& \{n^2(c-1)\}^{-1}\sum_{i=1}^{n}\sum_{h=1}^{H}\sum_{j<l}^{c}\{\varepsilon_i^2(X_{(h,j)}) + \varepsilon_i^2(X_{(h,l)})\}\textrm{ and }\\
I_5&\defby& \{n^2(c-1)\}^{-1}\sum_{i=1}^{n}\sum_{h=1}^{H}\sum_{j\neq l}^{c}\varepsilon_i(X_{(h,j)})\varepsilon_i(X_{(h,l)}).
\eeqrs

\noindent\textbf{Step 1.} We build the concentration inequality for $\wh\calS(X, Y)$. In this step, we consider $X$ is univariate random variable. 
We denote $\{(X_i,Y_{i})\}_{i=1}^{n}$ as independent sample following the same distribution as $(X,Y)$ throughout this step.

\noindent\textbf{Step 1.1.}  We derive the bound for $\pr(\abs{\wh{W} - W}>\eta)$. 
\beqrs
\pr(\abs{\wh{W} - W}>\eta) &\leq& 
\pr\big\{\abs{\wh{W} - E(\wh{W}\mid \{Y_{i}\}_{i=1}^{n})}>\eta/2\big\} \\
&&\hspace{3cm} + \pr\big\{\abs{E(\wh{W}\mid \{Y_{i}\}_{i=1}^{n}) - W}>\eta/2\big\}.
\eeqrs
We calculate the above two terms respectively.  
We have $\pr\big\{\abs{\wh{W} - E(\wh{W}\mid \{Y_{i}\}_{i=1}^{n})}>\eta/2\big\} = E\Big[\pr\big\{\abs{\wh{W} - E(\wh{W}\mid \{Y_{i}\}_{i=1}^{n})}>\eta/2\mid \{Y_{i}\}_{i=1}^{n}\big\}\Big]$.
Given the fact that $\wh{W}$ is bounded by $1/2$, using Hoeffding inequality, we have $\pr\big\{\abs{\wh{W} - E(\wh{W}\mid \{Y_{i}\}_{i=1}^{n})}>\eta/2\mid \{Y_{i}\}_{i=1}^{n}\big\}\leq 2\exp(-n\eta^2/2)$. 
This tail probability bound holds uniformly, thus
\beqr\label{equation:W1}
\pr\big\{\abs{\wh{W} - E(\wh{W}\mid \{Y_{i}\}_{i=1}^{n})}>\eta/2\big\} \leq 2\exp(-n\eta^2/2).
\eeqr
$E(\wh{W}\mid \{Y_{i}\}_{i=1}^{n})$ is U statistics with order $2$ and kernel bounded by $1/2$, we apply concentration inequality for U statistics \citep{hoeffding1963probability}, we have
\beqr\label{equation:W2}
 \pr\big\{\abs{E(\wh{W}\mid \{Y_{i}\}_{i=1}^{n}) - W}>\eta/2\big\}\leq 2\exp(-n\eta^2).
\eeqr
Combining the results of \eqref{equation:W1} and \eqref{equation:W2}, we have $\pr(\abs{\wh{W} - W}>\eta) \leq 4\exp(-n\eta^2/2)$.

\noindent\textbf{Step 1.2.} We derive the bound for $\pr(\abs{\wh\Lambda - \Lambda}>\eta)$. We have
\beqrs
\pr(\abs{\wh\Lambda - \Lambda}>\eta)\leq
\pr(\abs{I_1}>\eta/3) + \pr(\abs{2I_2}>\eta/3) + \pr(\abs{I_3-\Lambda}>\eta/3).
\eeqrs
Firstly, we consider $\pr(\abs{I_3-\Lambda}>\eta/3)$. We divide it into two parts.
\beqrs
\pr(\abs{I_3-\Lambda}>\eta/3)
\leq \pr(\abs{I_4-\Lambda}>\eta/6) 
+ \pr(\abs{I_5}>\eta/6).
\eeqrs
We consider the above two probability terms respectively.  
We have 
\beqrs
\pr(\abs{I_4-\Lambda}>\eta/6)  &\leq& \pr\big[\abs{I_4-E\{I_4\mid (X_i,Y_i)_{i=1}^{n}\}}>\eta/12\big]\\ 
&&\hspace{2cm}+ \pr\big[\abs{E\{I_4\mid (X_i,Y_i)_{i=1}^{n}\}-\Lambda}>\eta/12\big].
\eeqrs
Similarly as the derivation of \eqref{equation:W1}, we have $\pr\big\{\abs{I_4- E(I_4\mid (X_i,Y_i)_{i=1}^{n})}>\eta/12\big\}\leq 2\exp(-n\eta^2/72)$.
Additionally, $ E\{I_4\mid (X_i,Y_i)_{i=1}^{n}\}$ can be written as the sum of independent random variables, because sum of ordered sample equals to sum of non-ordered. We use Hoeffding inequality for random variables bounded by $1$ and get $\pr\big[\abs{E\{I_4\mid (X_i,Y_i)_{i=1}^{n}\}-\Lambda}>\eta/12\big]\leq 2\exp(-n\eta^2/288)$. Then, we get $\pr(\abs{I_4-\Lambda}>\eta/6)<4\exp(-n\eta^2/288)$.

We can also derive that  
$\pr(\abs{I_5}>\eta/6) \leq 
\pr\big[\abs{I_5 -E\{I_5\mid (X_i,Y_i)_{i=1}^{n}\}}>\eta/12\big] + \pr\big[\abs{ E\{I_5\mid (X_i,Y_i)_{i=1}^{n}\}}>\eta/12\big]$. Similarly as the derivation of \eqref{equation:W1}, we have $\pr\big[\abs{I_5 -E\{I_5\mid (X_i,Y_i)_{i=1}^{n}\}}>\eta/12\big]\leq 2\exp(-n\eta^2/288)$.
We have $\pr\big[\abs{ E\{I_5\mid (X_i,Y_i)_{i=1}^{n}\}}>\eta/12\big] = E\Big[\pr\big\{\abs{E\{I_5\mid (X_i,Y_i)_{i=1}^{n}\}}>\eta/12\mid \{X_i\}_{i=1}^{n}\big\}\Big]$.
For $h = 1,\ldots,H$, we denote
\beqrs 
I_{5,h}\defby\{c(c-1)\}^{-1}\sum_{j\neq l}^{c}E\big\{\varepsilon_i(X_{(h,j)})\varepsilon_i(X_{(h,l)})\mid (X_i,Y_i)_{i=1}^{n}\big\}.
\eeqrs
Conditioning on $\{X_i\}_{i=1}^{n}$, $I_{5,h}$ is U statistics using kernel bounded by $1$ with order 2.  
By applying concentration inequality for U statistics \citep{hoeffding1963probability}, we have
$\pr(\abs{I_{5,h}}>\eta\mid \{X_i\}_{i=1}^{n})\leq 2\exp(-c\eta^2/4)$. Therefore, $I_{5,h}$ is sub-Gaussian random variable with parameter $2/c$ conditioning on $\{X_i\}_{i=1}^{n}$. We have $I_5$ is the average of $I_{5,h}$ for $h = 1,\ldots,H$. And $I_{5,h}$ are independent. According to Azuma-Hoeffding  inequality, we know that  $I_{5}$ is sub-Gaussian random variable with parameter $2/(cH) = 2/n$. Therefore, we have
$\pr\big\{\abs{E\{I_5\mid (X_i,Y_i)_{i=1}^{n}\}}>\eta/12\mid \{X_i\}_{i=1}^{n}\big\}\leq 2\exp(-n\eta^2/576)$. 
This bound holds uniformly. Thus, we get $\pr\big[\abs{ E\{I_5\mid (X_i,Y_i)_{i=1}^{n}\}}>\eta/12\big]\leq 2\exp(-n\eta^2/576)$. 
Combining the two probability terms, we have
$\pr(\abs{I_5}>\eta/6)<4\exp(-n\eta^2/576)$.

Combining the results of $\pr(\abs{I_4-\Lambda}>\eta/6)$ and $\pr(\abs{I_5}>\eta/6)$, we have
\beqrs
\pr(\abs{I_3-\Lambda}>\eta/3)\leq O\{\exp(-n\eta^2/576)\}.
\eeqrs
Under Condition \ref{condition:sis:total_variation}-\ref{condition:sis:slice_order}, we apply Lemma \ref{lemma:sis:I1}, there exists $c_3>0$,
\beqrs
&&\pr(\abs{I_1}>\eta)\leq O[\exp\{-c_{3}n^{2-2\max(1/b,r) - 2\alpha}\eta^2\}]
\textrm{ and } \\
&&\pr(\abs{I_2}>\eta)\leq 
O[\exp\{-c_{3}n^{2-2\max(1/b,r) - 2\alpha}\eta^2\}].
\eeqrs 
Combine these three probabilities together and we can find $c_2>0$  that
\beqr\label{equation:concentration}
&&\pr\big\{\abs{\wh\calS(X,Y) - \calS(X,Y)}>\eta\big\}\nonumber \\
&=& \pr\big[\abs{(\wh{W}W)^{-1}\{\Lambda(\wh{W} - W) - W(\Lambda -\Lambda)\}}>\eta\big]\nonumber\\
&\leq& \pr\big\{\abs{(\wh{W}W)^{-1}\Lambda(\wh{W} - W)}>\eta/2\big\} 
+\pr\big\{\abs{\wh{W}^{-1}(\Lambda -\Lambda)}>\eta/2\big\} \nonumber\\
&\leq& O\{\exp(- c_{2}n^{\min\{1,2-2\max(1/b,r) - 2\alpha\}}\eta^2/c_{1}^{2})\}.
\eeqr
\noindent\textbf{Step 2.} We establish the sure screening property. From  \eqref{equation:concentration}, we have
\beqrs
&&\pr(\max_{1\leq k\leq p}\abs{\wh\omega_{k} - \omega_{k}}>c_{1}n^{-\gamma})\\
&\leq& \sum_{k=1}^{p}\pr(\abs{\wh\omega_{k} - \omega_{k}}>c_{1}n^{-\gamma})
\leq O\{p\exp(-c_{2}n^{\min\{1,2-2\max(1/b,r) - 2\alpha\}-2\gamma})\}.
\eeqrs

\noindent\textbf{Proof of Theorem \ref{theorem:sure_screening}:}
If $\calA\not\subseteq\wh\calA$, there exists $k\in\calA$ such that $\wh\omega_{k}<cn^{-\gamma}$. From Condition \ref{condition:sis_order}, we know $\abs{\wh\omega_{k} - \omega_{k}}>cn^{-\gamma}$ for some $k\in\calA$. Therefore, following the similar derivation in \cite{li2012feature}, we have
\beqrs
\pr(\calA\subseteq\wh\calA)
&\geq& 1 - \pr(\max_{k\in\calA }\abs{\wh\omega_{k} - \omega_{k}}>c_{1}n^{-\gamma}) \\
&=& 1 - \abs{\calA}\pr(\abs{\wh\omega_{k} - \omega_{k}}>c_{1}n^{-\gamma}) \\
&\geq& 1- O\{\abs{\calA}\exp(-c_{2}n^{\min\{1,2-2\max(1/b,r) - 2\alpha\}-2\gamma})\}.
\eeqrs

\noindent\textbf{Proof of Theorem \ref{theorem:rank_consis}:} 
Note that Condition \ref{condition:sis_order} holds, and the signals of inactive covariates satisfy $\omega_k=0$ for $k\in\calI$. Similar to the proof of Theorem \ref{theorem:sure_screening}, we have
\beqrs
\pr(\min_{k\in\calA}\wh\omega_k-\max_{k\in\calI}\wh\omega_k>0)
&\geq&
\pr(\min_{k\in\calA}\wh\omega_k> c_1n^{-\gamma}\ \mbox{and}\ \max_{k\in\calI}\wh\omega_k\leq c_1n^{-\gamma})\\
&\geq& 1 - \pr(\max_{1\leq k\leq p }\abs{\wh\omega_{k} - \omega_{k}}>c_{1}n^{-\gamma}) \\
&\geq&1 - p\pr(\abs{\wh\omega_{k} - \omega_{k}}>c_{1}n^{-\gamma})\\
&\geq&
1-O\big\{p\exp(-c_{2}n^{\min\{1,2-2\max(1/b,r) - 2\alpha\}-2\gamma})\big\}.
\eeqrs
\hfill$\fbox{}$


\section{Proof of Proposition \ref{prop:normality} }
Let $\gamma_1 = \{2\log n/(nW^2)\}^{1/2}>0$ and $\gamma_2 = (c/n)^{1/3}>0$.
\beqrs
\sup_{-\infty<t<\infty} \Big\lvert\pr\left[ \{n(c-1)\}^{1/2}\wh\calS(X,Y)/\sigma\leq t\right] - \Phi(t)\Big\rvert \leq P_{1} + P_{2} + P_{3},
\eeqrs
where
\beqrs
P_{1}&\defby&\sup_{-\infty<t<\infty} \Big\lvert\pr\left[ \{n(c-1)\}^{1/2}(\wh{W} - \wh{\Lambda})/(W\sigma)\leq t(1+\gamma_{1})\right] - \Phi(t)\Big\rvert,\\
P_{2}&\defby&\sup_{-\infty<t<\infty} \Big\lvert\pr\left[ \{n(c-1)\}^{1/2}(\wh{W} - \wh{\Lambda})/(W\sigma)\leq t(1-\gamma_{1})\right] - \Phi(t)\Big\rvert \textrm{ and }\\
P_{3}&\defby&\pr\left(\abs{\wh{W}/W - 1}>\gamma_{1}\right).
\eeqrs
We derive the bound for $P_{1}$ and omit the details for $P_2$, which follows similar paradigm.
TO simplify the $\wh{W} - \wh{\Lambda}$, we divide $\wh{W}$ into two parts, that $\wh{W} = \wh{W}_{1} - \wh{W}_{2}$, where
\beqrs
\wh{W}_1 &\defby& \{n^2(n-1)\}^{-1}\sum_{i=1}^{n}\sum_{j<l}^{n}
\{\nu_{i}^2(Y_{j}) + \nu_{i}^2(Y_{l})\} \textrm{ and }\\
\wh{W}_2 &\defby& \{n^2(n-1)\}^{-1}\sum_{i=1}^{n}\sum_{j\neq l}^{n}\nu_{i}(Y_{j})\nu_{i}(Y_{l}).
\eeqrs
Under independence,  we have $\wh{\Lambda} = I_{4} + I_{5}$. And it can be verified that $I_{4} = \wh{W}_{1}$. We further simplify $I_5$, and divide it into the main term $I_{6}$, and the remaining term $I_{7}$. 
\beqrs
I_5&=& \{n^2(c-1)\}^{-1}\sum_{i=1}^{n}\sum_{h=1}^{H}\sum_{j\neq l}^{c}\nu_i(Y_{(h,j)})\nu_i(Y_{(h,l)}) = I_{6} + I_{7},\textrm{ where }\\
I_{6}&\defby& \{n(c-1)\}^{-1}\sum_{h=1}^{H}\sum_{j\neq l}^{c}E_{T_{i}}\left\{\nu_i(Y_{(h,j)})\nu_i(Y_{(h,l)})\right\},
\eeqrs
and $I_{7} \defby I_{5} - I_{6}$. Given these facts, we have $\wh{W} - \wh{\Lambda} = I_{6} + I_{7} - \wh{W}_{2}$. Therefore, we can simplify $P_{1}$ and further divide it into three parts to derive the bound.
\beqrs
P_{1}&=&\sup_{-\infty<t<\infty} \Big\lvert\pr\left[ \{n(c-1)\}^{1/2}(I_{6} +  I_{7} - \wh{W}_{2})/(W\sigma)\leq t(1+\gamma_{1})\right] - \Phi(t)\Big\rvert\\
&\leq& P_{1,1} + P_{1,2} + P_{1,3},
\eeqrs
where
\beqrs
P_{1,1}&\defby&  \sup_{-\infty<t<\infty} \Big\lvert\pr\left[ \{n(c-1)\}^{1/2}I_{6} /(W\sigma)\leq t(1+\gamma_{1}) - \gamma_{2}\right] - \Phi(t)\Big\rvert,\\
P_{1,2}&\defby&  \sup_{-\infty<t<\infty} \Big\lvert\pr\left[ \{n(c-1)\}^{1/2}I_{6} /(W\sigma)\leq t(1+\gamma_{1}) + \gamma_{2}\right] - \Phi(t)\Big\rvert,\\
P_{1,3}&\defby& \pr\left(\abs{\{n(c-1)\}^{1/2}( I_{7} - \wh{W}_{2})}>\gamma_{2}\right).
\eeqrs
We first consider $P_{1,1}$ and divide it into two parts that $P_{1,1} = P_{1,1,1} + P_{1,1,2}$, where
\beqrs
P_{1,1,1} &\defby& \sup_{-\infty<t<\infty} \Big\lvert\pr\left[ \{n(c-1)\}^{1/2}I_{6} /(W\sigma)\leq t\right] - \Phi(t)\Big\rvert,\\
P_{1,1,2} &\defby& \sup_{-\infty<t<\infty}  \Big\lvert \Phi\{ t(1+\gamma_{1}) - \gamma_{2}\} - \Phi(t)\Big\rvert.
\eeqrs
To deal with $P_{1,1,1}$, we can adopt  Berry-Esseen inequality \citep{durrett2019probability}, by writing $\{n(c-1)\}^{1/2}I_{6}$ as the following formula.
\beqrs
\{n(c-1)\}^{1/2}I_6  = H^{-1/2}\sum_{h=1}^{H} I_{6,h},
~~ I_{6,h} \defby \{c(c-1)\}^{-1/2}\sum_{j\neq l}^{c}E_{T_{i}}\left\{\nu_i(Y_{(h,j)})\nu_i(Y_{(h,l)})\right\}.
\eeqrs
It is not difficult to verify that for $h=1,\ldots,H$, $E(I_{6,h}^2) = (W\sigma)^2\leq \infty$ and $E(\abs{I_{6,h}}^3)\leq \infty$. Given that $I_{6,h}$ are independent with each other, there exists $b_{1}>0$, such that $P_{1,1,1} \leq b_{1}(c/n)^{1/2}$. 

As for $P_{1,1,2}$, there exists $b_{2}>0$ such that
$P_{1,1,2}\leq b_{2}(t\gamma_{1} + \gamma_{2} )\max\big(\exp[-\{t(1+\gamma_1) - \gamma_2)\}^2/2], \exp(-t^2/2)\big)$.
From the definition, we have $\gamma_1<1/2$ and $\gamma_2<1/4$. Thus, when $\abs{t}\leq 2$, $P_{1,1,2}\leq 2b_{2}(\gamma_1 + \gamma_2)$. When $\abs{t}\geq 2$, $P_{1,1,2}\leq b_{2}(t\gamma_1 + \gamma_2)\exp(-t^2/8) \leq 4b_{2}(\gamma_1 + \gamma_2)$.

Combining $P_{1,1,1}$ and $P_{1,1,2}$, we have $P_{1,1} \leq b_{1}(c/n)^{1/2} + 4b_{2}(\gamma_{1} + \gamma_{2})$.
The bound for $P_{1,2}$ can be derived, following similar paradigm as $P_{1,1}$. 

Next, we consider $P_{1,3}$, which is bounded by $P_{1,3} \leq P_{1,3,1} + P_{1,3,2}$. Here, $P_{1,3,1}\defby\pr\left[\abs{\{n(c-1)\}^{1/2} I_{7}}>\gamma_{2}/2\right]$ and  $P_{1,3,2}\defby\pr\left[\abs{\{n(c-1)\}^{1/2} \wh{W}_{2}}>\gamma_{2}/2\right]$. 
By directly applying Markov inequality, there exists $b_{3} > 0$ and $b_{4}>0$, that  $P_{1,3,1}\leq b_{3}(1/n)/\gamma_{2}^2 = b_{3}(nc^2)^{-1/3}$ and
$P_{1,3,2}\leq b_{4}(c/n)/\gamma_{2}^2 = b_{4}(c/n)^{-1/3}$. Therefore, we can derive that $P_{1,3}\leq 2b_{4}(c/n)^{1/3}$.

When $n$ is sufficient large, $\gamma_{1}<\gamma_{2}$.
Combining $P_{1,1}$, $P_{1,2}$ and $P_{1,3}$, there exists $b_{5}>0$, such that $P_{1}\leq b_{5}(c/n)^{1/3}$.

The bound for $P_{2}$ can derived following similar derivation as $P_{1}$.To deal with $P_{3}$, we can make use of the result in step 1.1, proof of Theorem \ref{theorem:sis}. Thus, we have $P_{3}\leq 4\exp (-n\gamma_{1}^2W^2/2) = 4/n$.

Combining $P_{1}$, $P_{2}$ and $P_{3}$, we can derive the final result that
\beqrs
\sup_{-\infty<t<\infty} \Big\lvert\pr\left[ \{n(c-1)\}^{1/2}\wh\calS(X,Y)/\sigma\leq t\right] - \Phi(t)\Big\rvert \leq c_{0}(c/n)^{1/3}.
\eeqrs

\hfill$\fbox{}$

\section{Proof of Theorems \ref{theorem:SIT_BY}}

\noindent\textbf{Proof of Theorem \ref{theorem:SIT_BY:screening}:}
The key point is to determine the probability of the the event $\calA\subseteq\wh\calA(L^\star)\subseteq\wh\calA(L_q)$ where the first component has been considered by Theorem~\ref{theorem:sis}. Conditioning on the event $\calA\subseteq\wh\calA(L^\star)$, a sufficient condition for the second component is 
\beqrs
S(p)\frac{p\left\{1-\Phi\left(\{n(c-1)\}^{1/2}L^\star/\sigma \right)\right\}}{|\{k:\wh\omega_k\geq L^\star\}|\vee1}\leq q,
\eeqrs
according to the definition of $L_q$ in \eqref{equ:thrd_q}.The above inequality is fulfilled by 
\beqrs
p S(p)\left\{1-\Phi\left(c_1n^{1/2-\gamma}/\sigma \right)\right\}
\leq 
p \{1+\log(p)\} \exp(-c_1 n^{1-2\gamma}/\sigma^2)
\leq |\calA|q,
\eeqrs
where the first inequality is because of Gaussian tail and the second inequality holds according to \eqref{equ:p_n_cond}. Theorem~\ref{theorem:SIT_BY:screening} immediately stands by taking the unconditional probability according to Theorem~\ref{theorem:sis}.

\noindent\textbf{Proof of Theorem \ref{theorem:SIT_BY:FDR}:}
Denote the underlying distributions of $\{n(c-1)\}^{1/2}\wh\omega_k/\sigma$ are $\mathcal{N}_{n}$ for $k\in\calI$. The standard BY procedure is selecting $\wh\calA(\wt L_q)$ with
\beqrs
\wt L_q
&=& \inf \left\{t>0:p S(p)\frac{\pr(\mathcal{N}_n\geq \{n(c-1)\}^{1/2}t/\sigma)}{|\{k:\wh\omega_k\geq t\}|\vee1}\leq q \right\}\\
&:=& \inf \left\{t>0: \wh\FDP_{\mathcal{N}_n}(t)\leq q \right\}
\eeqrs
According to \citet[Theorem 1.3]{Benjamini2001},  
\beqr\label{equ:BY_FDR}
\FDR(\wh\calA(\wt L_q))\leq |\calI|q/p
\eeqr
regardless of the dependency among $\{\{n(c-1)\}^{1/2}\wh\omega_k/\sigma\}_{k=1}^p$. Since $\mathcal{N}_n$ is unknown in practice, a feasible approach is to replace $\mathcal{N}_n$ with its limiting distribution $\mathcal{N}_0$, the standard normal distribution. This approach is exactly our proposed SIT-BY procedure. 

The standard BY procedure and SIT-BY procedure differ in the two conservative estimators for $\FDP$, denoted as $\wh\FDP_{\mathcal{N}_0}(t)$ and $\wh\FDP_{\mathcal{N}_n}(t)$. Their difference can be quantified by
\beqrs
\wh\FDP_{\mathcal{N}_n}(t)
-
\wh\FDP_{\mathcal{N}_0}(t)
&=&
S(p)\frac{\pr(\mathcal{N}_n\geq \{n(c-1)\}^{1/2}t/\sigma)-1+\Phi(\{n(c-1)\}^{1/2}t/\sigma)}{\left(|\{k:\wh\omega_k\geq t\}|\vee1\right)/p}\\
&\defby&
D_{n,p}(t),
\eeqrs
Plugging $L_q$ and rearranging the above formula provide
\beqrs
\wh\FDP_{\mathcal{N}_n}(L_q)
=
\wh\FDP_{\mathcal{N}_0}(L_q) + D_{n,p}(L_q)
\leq q + D_{n,p}(L_q),
\eeqrs
where the inequality holds according to the definition of $L_q$ in \eqref{equ:thrd_q}. This indicates that $L_q$ is larger than the standard BY procedure threshold with the FDR level $q+D_{n,p}(L_q)$, i.e. $\widetilde{L}_{q+D_{n,p}(L_q)}$. Therefore, we have $\FDR(\wh \calA(L_q))\leq \FDR(\wh \calA(\widetilde{L}_{q+D_{n,p}(L_q)}))\leq |\calI|\{q+\mathbb{E}(D_{n,p}(L_q))\}/p$ {\color{red}according to \eqref{equ:BY_FDR}} and tower law by firstly conditioning on $D_{n,p}(L_q)$. For ease of notation, set $D_{n,p}(L_q)\defby D_{n,p}$ and $\wh\calA_q\defby\wh\calA(L_q)$. To bound $\mathbb{E}(D_{n,p}(L_q))$, we divide it into two parts:
\beqrs
\mathbb{E}(D_{n,p})
\leq 
\mathbb{E}\left(D_{n,p}\mid \calA\subseteq\wh\calA_q\right)\pr(\calA\subseteq\wh\calA_q)
+
\mathbb{E}\left(D_{n,p}\mid \calA\not\subseteq\wh\calA_q\right)
\pr(\calA\not\subseteq\wh\calA_q).
\eeqrs
For the first part, Proposition~\ref{prop:normality} suggests 
\beqrs
\mathbb{E}\left(D_{n,p}\mid \calA\subseteq\wh\calA_q\right)\pr(\calA\subseteq\wh\calA_q)
\leq c_0 S(p) \frac{ (c/n)^{1/3}}{|\calA|/p}
\leq c_0(c/n)^{1/3}\frac{p\{1+\log(p)\}}{|\calA|}.
\eeqrs
For the second part, we have
\beqrs
& &\mathbb{E}\left(D_{n,p}\mid \calA\not\subseteq\wh\calA_q\right)
\pr(\calA\not\subseteq\wh\calA_q)\\
&\leq&
O\big\{(c/n)^{1/3} p S(p) |\calA| \exp(-c_{2}n^{\min\{1,2-2\max(1/b,r) - 2\alpha\}-2\gamma})\big\}\\
&\leq&
O\big\{ (c/n)^{1/3}p^2 \log(p) \exp(-c_{2}n^{\min\{1,2-2\max(1/b,r) - 2\alpha\}-2\gamma})\big\}.
\eeqrs
where the first inequality is because of Proposition~\ref{prop:normality} and Theorem~\ref{theorem:SIT_BY:screening}, and the second inequality is due to $|\calA|\leq p$.
These two parts show that $\FDR(\wh \calA(L_q))\leq |\calI|\{q+ O(c_{n,p,|\calA|})\}/p$, where
\beqrs
c_{n,p,|\calA|}=p\log(p)(c/n)^{1/3}/|\calA|
+
(c/n)^{1/3} p^2 \log(p) \exp(-c_{2}n^{\min\{1,2-2\max(1/b,r) - 2\alpha\}-2\gamma}).
\eeqrs
Noticing that the convergence is dominated by the first part, we have $c_{n,p,|\calA|}\rightarrow 0$ as $n\rightarrow\infty$ under the condition $p\log(p)/|\calA|=o\{(n/c)^{1/3}\}$. 

Finally, we obtain $\lim_{(n,p)\rightarrow\infty}\FDR(\wh\calA(L_q))\leq q$.\hfill$\fbox{}$


\section{Technical Lemmas}
\label{sec:technical_lemmas}

{\lemm\label{lemma:ranktail}
	Suppose $Z_1, \ldots, Z_n$ are an i.i.d. sample and $r$ is a positive constant. Let $Z_{(i)}$ be the $i$-th order statistic. If for any $\eta>0$ and $x>0$, $x^r pr(\abs{Z}>x\eta) \leq \exp(-\eta^2)$, then
	$\pr\{n^{-1/r}(\abs{Z_{(n)}} + \abs{Z_{(1)}})>\eta\}\leq 2\exp(-\eta^2/4)$.}

\noindent\textbf{Proof of Lemma \ref{lemma:ranktail}:}  By definition, we have
\beqrs
\pr\{n^{-1/r}Z_{(n)} >\eta/2\} &=& 1 - \pr(Z_{(n)}\leq n^{1/r}\eta/2)\\
&=& 1 - \{1-\pr(Z> n^{1/r}\eta/2)\}^{n}\\
&\leq& n\pr(Z> n^{1/r}\eta/2)\leq \exp(-\eta^2/4).
\eeqrs
We can similarly prove that
$\pr\{n^{-1/r}Z_{(1)} <-\eta/2\} \leq \exp(-\eta^2/4)$. And we complete the proof by adding these two probabilities.

\hfill$\fbox{}$

{\lemm\label{lemma:sis:I1}
	Under Condition \ref{condition:sis:total_variation}-\ref{condition:sis:slice_order},  there exists $c_3>0$,
	\beqrs
	&&\pr(\abs{I_1}>\eta)\leq O[\exp\{-c_{3}n^{2-2\max(1/b,r) - 2\alpha}\eta^2\}]
	\textrm{ and } \\
	&&\pr(\abs{I_2}>\eta)\leq 
	O[\exp\{-c_{3}n^{2-2\max(1/b,r) - 2\alpha}\eta^2\}].
	\eeqrs
}
\noindent\textbf{Proof of Lemma \ref{lemma:sis:I1}:} 
We prove the probability bound for $I_2$ and omit the proof for $I_1$, which can be proved similarly. 
Following the deviation of equation (S.5.1) in the Supplement for ``Sliced Independence Test'', 
\beqrs
\pr(\abs{I_2}> \eta) 
&\leq&
\pr\left\{2c/n^2\sum_{i = 1}^{n} \sum_{j = 1}^{n - 1}\abs{s_i(X_{(j + 1)}) - s_i(X_{(j)})}>\eta\right\}\\
&\leq& \pr\left\{\sup_{T_{i}}2c/n\sum_{j = 1}^{n - 1}\abs{s_i(X_{(j + 1)}) - s_i(X_{(j)})}>\eta\right\}
\eeqrs
If $X$ has a bounded support, by  \ref{condition:sis:total_variation},  
\beqrs
\pr(\abs{I_2}> \eta) &\leq& \pr\left\{\sup_{\Pi_n(B),T_{i}}n^{-r}\sum_{j = 1}^{n - 1}\abs{s_i(X_{(j + 1)}) - s_i(X_{(j)})}>\eta n^{1-r}/(2c)\right\}\\
&\leq& \exp\left\{-n^{2-2r}\eta^2/(4c^2)\right\}.
\eeqrs
If the support of $X$ is unbounded, it suffices to show that for $\delta\in(0, 1/2)$,
\beqr\label{lemma:sis:T2:eq1}
&&\pr\left\{\sup_{\Pi_n(B),T_{i}}n^{-r}\sum_{j = [n\delta]}^{[n(1 - \delta)]}\abs{s_i(X_{(j + 1)}) - s_i(X_{(j)})}>\eta n^{1-r}/(4c)\right\}\nonumber\\
&\leq& 3\exp\left\{-n^{2-2r}\eta^2/(16c^2)\right\}.
\eeqr
and, 
\beqr\label{lemma:sis:T2:eq2}
&& \pr\left\{n^{-1/b}\sum_{j = 1}^{[n\delta]} \abs{s_i(X_{(j + 1)}) - s_i(X_{(j)})} > \eta n^{1-1/b}/(8c)\right\} +\nonumber\\ 
&&  \pr\left\{n^{-1/b}\sum_{j = [n(1 - \delta)]}^{n - 1} \abs{s_i(X_{(j + 1)}) - s_i(X_{(j)})} > \eta n^{1-1/b}/(8c)\right\} \nonumber\\ 
&\leq&
4\exp\left\{-n^{2-2/b}\eta^2/(64c^2)\right\}
\eeqr
We shall show \eqref{lemma:sis:T2:eq1} and \eqref{lemma:sis:T2:eq2} in the following two steps.

\noindent\textbf{Step 1.} We aim to show \eqref{lemma:sis:T2:eq1} holds. In the proof of Lemma \ref{lemma:I2}, we denote $A_n = 1\{X_{([n\delta])}> G^{\leftarrow}(\beta)\}$ and $B_n = 1\{X_{([n(1 - \delta)])} <  G^{\leftarrow}( 1 - \beta)\}$ for $0< \beta < \delta$. 
Here we choose $\beta$ that satisfying both the following inequalities.
\beqrs
&&\pr\{X_{([n\delta])}\leq G^{\leftarrow}(\beta)\}\leq \exp\{-n^{2-2r}\eta^2/(16c^2)\} \textrm{ and }\\
&&\pr\{X_{([n(1-\delta)])}\geq G^{\leftarrow}(1-\beta)\}\leq \exp\{-n^{2-2r}\eta^2/(16c^2)\}.
\eeqrs
Then, \eqref{lemma:sis:T2:eq1} can be established given the following facts,
\beqrs
&&\pr\left\{\sup_{\Pi_n(B),T_{i}}n^{-r}\sum_{j = [n\delta]}^{[n(1 - \delta)]}\abs{s_i(X_{(j + 1)}) - s_i(X_{(j)})}A_{n}B_{n}>\eta n^{1-r}/(4c)\right\}\nonumber\\
&\leq& \exp\left\{-n^{2-2r}\eta^2/(16c^2)\right\},
\eeqrs
which, in turns, follows from Condition  \ref{condition:sis:total_variation} with $r$.

\noindent\textbf{Step 2.} We aim to show (\ref{lemma:sis:T2:eq2}).
In the proof of Lemma \ref{lemma:I2}, we denote  $C_n = 1(X_{([n\delta])} < -B_0)$. We choose $\delta$ that satisfying that $\pr\{X_{([n\delta])} \geq -B_0\}\leq \exp\{-n^{2-2/b}\eta^2/(64c^2)\}$.
By the non-expansive condition in  \ref{condition:sis:total_variation}, we have
\beqrs
\sum_{j = 1}^{[n\delta]} \abs{s_i(X_{(j + 1)}) - s_i(X_{(j)})}C_n & \leq & \sum_{j = 1}^{[n\delta]} \abs{M(X_{(j + 1)}) - M(X_{(j)})} \\
& = & \abs{M(X_{(1)}) - M(X_{([n\delta])})},
\eeqrs
Using the above fact, we can build the probability bound that
\beqrs
&&\pr\left\{n^{-1/b}\sum_{j = 1}^{[n\delta]} \abs{s_i(X_{(j + 1)}) - s_i(X_{(j)})}C_{n} > \eta n^{1-1/b}/(8c)\right\} \\
&\leq&\pr\left\{ n^{-1/b}\abs{M(X_{(1)}) - M(X_{([n\delta])})}>\eta n^{1-1/b}/(8c)\right\}\\
&\leq& \pr\left\{ n^{-1/b}\abs{M(X_{(1)})} + \abs{M(X_{([n])})}>\eta n^{1-1/b}/(8c)\right\}\\
&\leq& \exp\left\{-n^{2-2/b}\eta^2/(64c^2)\right\},
\eeqrs 
The last inequality holds under Condition \ref{condition:sis:total_variation} for Lemma \ref{lemma:ranktail}. The other tail can be dealt with similarly and we complete the proof for \eqref{lemma:sis:T2:eq2}.
Combining steps 1-2, there exists $c_3$ that
\beqrs
\pr(\abs{I_2}> \eta) &\leq& 4\exp\left\{-n^{2-2\max\{1/b,r\}}\eta^2/(64c^2)\right\}\\
&\leq& O[\exp\{-c_{3}n^{2-2\max\{1/b,r\} - 2\alpha}\eta^2\}]
\eeqrs
The last inequality holds for Condition \ref{condition:sis:slice_order}. 
We complete the proof of Lemma \ref{lemma:sis:I1}.\hfill$\fbox{}$

\end{document}